\documentclass[a4paper,12pt]{article}
\usepackage{fullpage}
\usepackage{color}
\definecolor{lacre}{rgb}{0.67,0,0}
\usepackage[centertags]{amsmath}
\usepackage{amscd,amsfonts,amssymb,latexsym}
\usepackage[dvips]{graphicx}
\usepackage{theorem}
\usepackage{enumerate}
\usepackage[toc,page]{appendix}
\usepackage{textcomp}
\usepackage{flafter}

\newcommand{\al}{\alpha}
\newcommand{\ra}{\rightarrow}
\newcommand{\e}{\epsilon}
\newcommand{\dd}{\partial}
\newcommand{\eps}{\epsilon}
\newcommand{\ordre}{\mathcal{O}}

\newcommand{\m}{\mu}
\newcommand{\p}{\psi}
\newcommand{\n}{\nabla}
\newcommand{\del}{\nabla}
\newcommand{\ki}{\chi}
\newcommand{\beq}{\begin{equation}}
\newcommand{\eeq}{\end{equation}}
\newcommand{\beqas}{\begin{eqnarray*}}
\newcommand{\eeqas}{\end{eqnarray*}}
\newcommand{\beqa}{\begin{eqnarray}}
\newcommand{\eeqa}{\end{eqnarray}}
\newcommand{\g}{G}
\theoremstyle{plain}

\author{M. Aguareles \and  S.J. Chapman}
\title{Motion of spiral waves in the Complex Ginzburg-Landau equation}
\begin{document}
\maketitle

\begin{abstract}
Solutions of the general cubic complex Ginzburg-Landau equation
comprising multiple spiral waves are considered. For parameters close
to the vortex limit, and for a system of
spiral waves with  well-separated centres, laws of motion of 
 the centres are found which vary  depending on the order of magnitude
 of the separation of the centres. 
In particular, the direction of the interaction
changes from along the line of centres to perpendicular to the line of
centres as the separation increases, with the strength of the
interaction
algebraic at small separations and exponentially small at large
separations.  The corresponding asymptotic wavenumber and frequency
are determined. These depend on
the positions of the 
centres of the spirals,  and so evolve slowly as the spirals move.
\end{abstract}
\section{Introduction}
The complex Ginzburg-Landau equation is one of the most-studied
nonlinear models in physics. It describes on a qualitative level,
and in many important cases on a quantitative level, a great number of
phenomena, from nonlinear waves to second-order phase transitions, including
superconductivity, superfluidity, Bose-Einstein condensation, liquid 
crystals, and string theory \cite{aranson02}.

The equation arises as the amplitude equation in the vicinity of a
Hopf bifurcation in spatially extended systems, and is therefore
generic for active media displaying wave patterns. The simplest
examples of such media are chemical oscillations such as the famous
Belousov-Zhabotinsky reaction. More complex examples include
thermal convection of binary fluids \cite{AransonRef6} and transverse
patterns of high intensity light \cite{moloney90}. 

The general cubic complex Ginzburg-Landau equation is given by 
\begin{equation}
\frac{\partial \Psi}{\partial t} =
\Psi-(1+i a) \ |\Psi|^2 \Psi+(1+i b) \n^2 \Psi,
\label{CGL0}
\end{equation}
where $a$ and $b$ are real parameters and the complex field $\Psi$
represents the amplitude and phase of the modulations of the
oscillatory pattern.  

Of particular interest are ``defect'' solutions of (\ref{CGL0}). These
are topologically stable solutions in which $\Psi$ has a single zero,
around which the phase of $\Psi$ varies by a non-zero integer multiple
of $2 \pi$. When $a=b$ these solutions are known as ``vortices'', and
the constant phase lines are rays emanating from the zero. When $a
\not = b$ the defect solutions are known as ``spirals'', with the
constant phase lines behaving as rotating Archimedean spirals except
in the immediate vicinity of the core. 

It is often convenient to factor out the rotation of the spiral, by writing
\[
\Psi =e^{-i\omega  t}\sqrt{\frac{1+\omega b}{1+ab}}\,\psi,
\qquad t = \frac{t'}{1+\omega b}, \qquad {\bf x} =
\sqrt{\frac{1+b^2}{1+b\omega}} \, {\bf x}'.
\] 
This gives, on dropping the primes,
\begin{equation}
(1-i\,b)\frac{\partial \psi}{\partial t} = \n^2 \psi+(1-|\psi|^2)\psi+iq\psi(1-k^2-|\psi|^2) ,
\label{CGL}
\end{equation}
where
\begin{equation*}
q=\frac{a-b}{1+b a}, \qquad q(1-k^2)=\frac{\omega-b}{1+b \omega};
\end{equation*}
rotating single spiral waves are now stationary solutions of \eqref{CGL}. The
constant $k$ is known as the \emph{asymptotic wavenumber}, since it is
easily shown that at infinity $\arg(\psi) \sim n\phi \pm k r$. An
important property of  spiral wave solutions is that the asymptotic
wavenumber $k$ is not a free parameter, but is uniquely determined by
$q$ \cite{hagan82}. Physical systems corresponding to (\ref{CGL}) generally contain not
one but many defects. Such complex patterns may be understood in terms of the position of these defects. Thus if the motion
of the defects can be determined, much of the dynamics of the solution
can be understood. 

Defect solutions of \eqref{CGL} behave very differently depending on
whether  $q=0$ (corresponding to $a=b$) or $q\not = 0$ (corresponding to $a \not =
b$). When  $q=0$ the wavenumber $k=0$, and  a great amount is
known about the solutions to \eqref{CGL}. In particular, in a seminal
work,  Neu \cite{neu90} analysed a system of $n$ vortices
asymptotically in the limit in which the separation of vortices is
much greater than the core radius, using the theory of matched
asymptotic expansions. By approximating the solution using near-field
or ``inner'' expansions in the vicinity of each vortex core and
matching these to a far-field or ``outer'' expansion away from vortex
cores, Neu derived a law of motion for each vortex in terms
of the positions of the others, thus reducing \eqref{CGL} to the
solution of $2n$ ordinary differential equations (for the $x$- and
$y$-coordinates of each vortex). The interaction between defects in
this case is long-range, essentially decaying like $r^{-1}$ for large
$r$. Neu's analysis has become the template for the analysis of the
motion of a system of singularities in many equations, including more
detailed models of superconductivity \cite{peres93}, \cite{chapman2}, \cite{chapman3}.  As we shall show, the key property of \eqref{CGL} that facilitated Neu's analysis is the
linearity of the far-field equations. Thus the contribution from many vortices in the far field can be obtained by a simple linear superposition, and the motion of vortices
is determined through the interaction of this far-field with the
individual core solutions.  

When $q>0$ the wavenumber $k>0$, and the situation is much more
complicated, even for a single defect. Hagan \cite{hagan82} studied
single spiral wave solutions of \eqref{CGL} and demonstrated that the
asymptotic wavenumber $k$ is uniquely determined by $q$. Using
perturbation techniques he found asymptotic expressions for $\psi$ and
the asymptotic wavenumber $k$ as a function of $q$ and the winding number of
the spiral $n$. For small values of $q$ a single defect has a
multi-layer structure, with the solution comprising inner, outer and
far-field regions in which different 
approximations hold. The transition from the outer region to the
far-field 
 occurs exponentially far (in $q$) from the centre (at what we shall
 term the outer core 
radius), and the asymptotic wavenumber $k$ is 
correspondingly exponentially small in $q$. This outer core radius is
the radius at which the level phase lines switch from being
essentially radial to essentially azimuthal. Thus for non-zero $q$ there is a new lengthscale in the problem, with
each spiral core having two lengthscales. In studying the motion of
spirals it is no longer enough to say that they are well-separated by
comparison to the core radius; now it must be determined whether the
separation is large compared to the inner core, but small compared to
the outer, or whether the separation is large compared to the outer 
core radius so that the interaction is truly far-field. When the
separation lies between the inner and outer core radii the interaction
is algebraic, but when the separation is large compared to the outer
core radius the interaction of defects decays exponentially.  

The fact that the outer equation for the phase of $\psi$ is nonlinear,
so that the contributions from multiple defects may not simply be
added, along with the exponential scaling of the outer variable, explains the
difficulty in applying Neu's techniques to the general case of
non-zero $q$. Thus, despite much work and some 
partial results \cite{aranson93,pismen92}, the interaction of defects
in the case of non-zero $q$ was  not completely
understood. However, recently in \cite{ACW} a set of laws of motion for $N$
 spirals with unit winding number was derived systematically in the
 limit $0 \leq q \ll 1$. The aim of 
 the present work is to give the details of that calculation. 

We start in Section \ref{q=0} with
reviewing the general asymptotic scheme that determines the spirals'
mobility when $q=0$. In Section \ref{sim} we inspect the equilibrium
solutions to \eqref{CGL} for $q>0$ formed by a single spiral,
highlighting the existence of two distinguished outer regions  where either the
azimuthal or the radial components of the phase function dominate. 
 These two sections serve as
a template for the analysis of multiple-spirals patterns, which is
performed in Sections \ref{canonical} and \ref{middle}. In Section
\ref{canonical} we derive a law of motion for spirals which are
separated by distances comparable to the outer core radius; since this
is a distinguished limit we refer to it as the \emph{canonical}
separation.
 The
interaction at the canonical separation is found to be exponentially
small and it takes place in the direction perpendicular to the line
of centres of the spirals.  In Section \ref{middle} spirals are
assumed to be separated 
by distances lying between the inner and outer core radii; 
we  denote this as the \emph{near-field} separation. In this case
the interaction becomes algebraic with a component along the line of
the centres. 

Changing the separation of the spirals while keeping the parameter $q$
fixed is equivalent to varying $q$ at fixed separation. We will see
that the near-field
separation is needed to  interpolate
between the canonical separation and the case $q=0$. 

\section{Interaction of vortices in the Ginzburg-Landau 
equation with real   coefficients} 
\label{q=0}
Without lost of generality and to simplify the calculations we
consider equation \eqref{CGL} with $b=0$. We return briefly to the
case of general $b$ in the appendix. For $q=0$ this reads
\begin{equation}
\Psi_t = \Psi(1 -|\Psi|^2) + \nabla^2\Psi \label{3},
\end{equation}
which by writing $\Psi = fe^{i\ki}$ with $f$ and $\chi$ real and
separating real and imaginary parts we may write as
\begin{eqnarray}
f_t &=& \nabla^2 f -f|\n\ki|^2+(1-f^2)f,\label{4}\\
f\ki_t &=& 2\n\ki\n f+f\nabla^2\ki.\label{5}
\end{eqnarray}
We wish to determine the law of motion for well separated vortices
following \cite{neu90}. We
assume that the separation is $O(\eps^{-1})$, with $\eps \ll 1$. 
This leads to an ``outer region'', in which ${\bf x}$ is scaled with
$\eps^{-1}$, and an ``inner region'' in the vicinity of each
vortex. Matching the asymptotic expansions of the solutions in each of
these regions leads to the law of motion for each vortex centre.

\subsection{Outer region}
In the outer region we rescale ${\bf x}$ and $t$ by setting ${\bf X} =
\e {\bf x}$ and $T =\e^2\m t$; here $\m$ is a small
parameter (the timescale for vortex motion) which will be determined
later. We will find that $\mu$ is logarithmic in $\eps$.
With this rescaling (\ref{4}), (\ref{5}) read
\begin{eqnarray}
\e^2\m f_T &=& \e^2(\nabla^2 f -f|\n\ki|^2)+(1-f^2)f,\\
\m f\ki_T &=& 2\n\ki\cdot\n f+f\nabla^2\ki.
\end{eqnarray}
Expanding in powers of $\eps$ as 
\beqas
f &\sim& f_0+\e^2 f_1+\cdots,\\
\ki &\sim& \ki_0 +\e^2 \ki_1+\cdots
\eeqas
 we find 
\begin{eqnarray}
f_0 &=& 1,\\
\m\frac{\dd \ki_{0}}{\dd T} &=& \nabla^2\ki_0 \label{9}
\end{eqnarray}
Now expanding $\chi_0$ for small $\mu$ as 
\[
\ki_0 \sim \ki_{00}+\m\ki_{01}+\cdots
\]
and substituting into (\ref{9}) we obtain at leading order
\[
\nabla^2\ki_{00}=0,
\]
with solution
\[ \ki_{00} =
  \sum_{j=1}^Nn_j \phi_j,\]
where $\phi_j$ is the polar angle measured from the centre of the
$j$\,th vortex ${\bf X}_j$, and $n_j$ is the winding number (or
degree) of the
$j$\,th vortex. At the next order in $\mu$ we find
\[
\nabla^2\ki_{01}=\frac{\dd \ki_{00}}{\dd T},\]
with solution
\[
\ki_{01}=-\frac{1}{2}\sum_{j=1}^N n_j R_j\log R_j \, {\bf e}_{\phi
  j}\cdot\frac{d{\bf X}_j}{dT},
\]
where $R_j$ is the  distance from the $j$\,th vortex.
Continuing to $O(\mu^2)$  we find
\[
\ki_{02} = \frac{1}{8}\sum_{j=1}^N n_j R_j^2\log R_j \left({\bf e}_{\phi
  j}\cdot\frac{d{\bf X}_j}{dT}\right)\left({\bf e}_{r
  j}\cdot\frac{d{\bf X}_j}{dT}\right).
\]
In general we find that $\chi_{0m}$ is $O(R^m \log R)$ as $R \ra 0$.

\subsection{Inner Region}
We rescale near the $\ell$\,th vortex by setting ${\bf X} = {\bf X}_{\ell} +
\eps {\bf x}$ to give
\begin{eqnarray}
\e\m \left(
\e f_T-\frac{d{\bf X}_{\ell}}{dT}\cdot \n f
\right) &=& \nabla^2 f -f|\n\ki|^2+(1-f^2)f,\\
\e\m \left(
\e f\ki_T-f\frac{d{\bf X}_{\ell}}{dT}\cdot \n \ki
\right) &=& 2\n\ki\n f+f\nabla^2\ki,
\end{eqnarray}
or equivalently
\begin{equation}
\e\m\left(\e\Psi_T -\frac{d{\bf X}_{\ell}}{dT}\cdot \n\Psi\right)=
\Psi(1 -|\Psi|^2) + \nabla^2\Psi. 
\end{equation}
Expanding in powers of $\eps$ as 
$\Psi \sim \Psi_0+\e\Psi_1+\ldots$ we find at leading order
\beq
\nabla^2\Psi_0+\Psi_0(1 -|\Psi_0|^2) =0.\label{loinnereqn}
\eeq
This is just the equation for a single static vortex, 
with solution
\begin{equation}
\Psi_0 = f_0 \,e^{i\chi_0}= f_0(r)e^{i(n_{\ell}\phi +C(T))},
\label{loinner}
\end{equation}
where
\[
\frac{d^2 f_0}{dr^2} + \frac{1}{r}\frac{d f_0}{dr} - f_0\frac{n_{\ell}^2}{r^2} +
f_0 - f_0^3 = 0,
\]
\[ f(0) = 0, \qquad f \ra 1 \mbox{ as } r \ra \infty,\]
and  $C(T)$ is determined by matching. It is well known  that this
equation has a unique increasing monotone solution \cite{greenberg80}.

Continuing with the expansion
we find at first order in $\eps$ that
\begin{equation}
-\m\n\Psi_0\cdot\frac{d{\bf X}_{\ell}}{dT} = \n^2\Psi_1+\Psi_1(1
 -|\Psi_0|^2)-\Psi_0(\Psi_0\Psi_1^*+\Psi_0^*\Psi_1).
\label{firstinner0}
\end{equation}

\subsubsection{Solvability Condition}
\label{q=0solv}
Since the linear operator on the right-hand side of equation
\eqref{firstinner0} is 
self-adjoint, and the homogeneous version of \eqref{firstinner0} is
satisfied by the partial derivatives of $\Psi_0$ (as can be seen by
differentiating \eqref{loinnereqn}), by the Fredholm Alternative there is
a solvability condition on (\ref{firstinner0}) which we can write as
\begin{equation*}
-\int_D \Re\left\{\m\left(\n\Psi_0\cdot\frac{d{\bf
        X}_{\ell}}{dT}\right)\left(\n\Psi_0\cdot 
  {\bf d}\right)^*\right\} \, dD 
= \int_{\partial D} \Re\left\{(\n\Psi_0\cdot
{\bf d})\frac{\partial\Psi_1^*}{\partial
  r}-\Psi_1^*\frac{\partial(\n\Psi_0\cdot {\bf d})}{\partial
  r}\right\}\, dl
\end{equation*}
where ${\bf d}$ is an arbitrary constant vector and $D$ is an arbitrary region
in the plane. Taking $D$ to be a ball of radius $r$ we find, after
some calculations, 
\begin{equation}
-\m\pi\left(\frac{d{\bf X}_{\ell}}{dT}\cdot {\bf d}\right)\int_0^r\left( s
  (f'_0)^2+n_k^2\frac{f_0^2}{s}\right)\,ds = n_k
 \int_0^{2\pi}\left(\frac{\partial\chi_1}{\partial
 r}+\frac{\chi_1}{r}\right){\bf e}_{\phi }\cdot {\bf d} \, d\phi
\label{solcond0}
\end{equation}
where $\chi \sim \chi_0 + \eps \chi_1 + \cdots$. After matching with
the outer region to determine $\chi_1$, equation
\eqref{solcond0} will determine vortex  velocity $d{\bf
    X}_{\ell}/dT$.  

\subsection{Asymptotic matching}
\subsubsection{Inner limit of the outer}
We express the leading-order (in $\eps$) outer $\ki_0$ in terms of the 
inner variable ${\bf x}$ given by  
${\bf X} = {\bf X}_{\ell} + \eps {\bf x}$ so that $R_{\ell} = \eps r$, 
$\phi_{\ell}
= \phi$. Then, Taylor-expanding $\chi_0$, 
\begin{eqnarray}
\ki_0 &\sim& \ki_{00}+\m\ki_{01}+\ldots\nonumber\\ 
&\sim& n_{\ell}\phi+G({\bf X}_{\ell})-n_{\ell}\frac{\e r}{2}\log\e r\,
{\bf e}_{\phi 
  }\cdot\frac{d{\bf X}_{\ell}}{dT}\m+\e\n G({\bf X}_{\ell})\cdot {\bf
    x}+\ordre(\e^2), 
\label{2ti1to}
\end{eqnarray}
where 
\begin{eqnarray*}
G({\bf X}) &=& \sum_{j\neq {\ell}}
n_j\phi_j-\m\sum_{j\neq k} \frac{n_j}{2}|{\bf X}-{\bf X}_j|\log|{\bf X}-{\bf X}_j|\frac{d{\bf X}_j}{dT}\cdot {\bf e}_{\phi
  j}\\ 
&&\mbox{ }+\mu^2 \sum_{j\neq {\ell}}\frac{n_j}{8}|{\bf X}-{\bf X}_j|^2\log |{\bf X}-{\bf X}_j| \left({\bf e}_{\phi
  j}\cdot\frac{d{\bf X}_j}{dT}\right)\left({\bf e}_{r
  j}\cdot\frac{d{\bf X}_j}{dT}\right) +\ordre(\m^3).
\end{eqnarray*}

\subsubsection{Outer limit of the inner}
The leading-order phase in the inner region is
\[ \chi_0 = n_ {\ell}\phi + C(T).\]
This matches with the first term in (\ref{2ti1to}) providing $C(T) = G({\bf X}_ {\ell})$.

The two-term inner expansion for the phase is $\chi_0 + \eps \chi_1$.
This should match with the two-term inner expansion of the outer
(\ref{2ti1to}).
Since $\mu = O(1/\log 1/\eps)$ and all logarithmic terms need to be
matched at the same time, to perform this matching we need to
take the full $\mu$ 
expansion of both terms. Fortunately only the expansion of $G({\bf X}_
{\ell})$ and 
$\del G({\bf X}_ {\ell})$
involve infinitely many terms in $\mu$, and these are evaluated at
${\bf X}_ {\ell}$ and are therefore independent of ${\bf x}$. 
We see that the expansions match if
\beq
 \chi_1 \sim -n_{\ell}\frac{ r}{2}\log\e r\, {\bf e}_{\phi
  }\cdot\frac{d{\bf X}_{\ell}}{dT}\m+\n G({\bf X}_{\ell})\cdot {\bf x} \label{chi1}
\eeq
as $r \ra \infty$, which is good to all orders in $\mu$.

\subsection{Law of motion}
We can now use the matching condition (\ref{chi1}) in the solvability
condition (\ref{solcond0}) to find a law of motion for each vortex.

As $r \ra \infty$ the left-hand side of (\ref{solcond0}) is 
\begin{equation}
-\m\pi(n_{\ell}^2\log r +a)\, {\bf d}\cdot\frac{d{\bf X}_{\ell}}{dT}
\end{equation}
where 
\begin{equation*}
a = \lim_{r\to\infty} \left[\int_0^r \left(s
  (f'_0)^2+n_{\ell}^2\frac{f_0^2}{s}\right)\,ds - \log r\right], 
\end{equation*}
is a constant independent of $\m$ and $\e$.
Using  (\ref{chi1}) in the right-hand side of (\ref{solcond0}) gives
\begin{equation}
-\m\pi (1/2+\log r)n_{\ell}^2\, {\bf
  d}\cdot\frac{d{\bf X}_{\ell}}{dT}
-\m n_{\ell}^2\log\e\pi {\bf
  d}\cdot\frac{d{\bf X}_{\ell}}{dT}+ 2\pi {\bf d} \cdot \del  G^{\perp}({\bf X}_{\ell}), 
\end{equation}
where $^{\perp}$ represents rotation of a vector by $\pi/2$. 
Since $\m$ is a small
constant, we find that the only way to make the left and right hand
sides balance is by taking $\m = 1/\log(1/\e)$, giving
\begin{equation}
-a\m \,{\bf d}\cdot\frac{d{\bf X}_{\ell}}{dT} =n_{\ell}^2
\, {\bf d}\cdot\frac{d{\bf X}_{\ell}}{dT}-\m\frac{n_{\ell}^2}{2}
\, {\bf   d}\cdot\frac{d{\bf X}_{\ell}}{dT}+2n_{\ell}
\,{\bf d} \cdot \del  G^{\perp}({\bf X}_{\ell}).
\end{equation}
Since ${\bf d}$ is arbitrary this can be rearranged to give the law of motion
\begin{equation}
\frac{d{\bf X}_{\ell}}{dT} = -\frac{2n_{\ell}\n G^{\perp}({\bf X}_{\ell})}{n_{\ell}^2+\m(a-n_{\ell}^2/2)}+\ordre(\e)
\label{eqLaw}
\end{equation}
where 
\begin{eqnarray*}
\n G({\bf X}_{\ell}) &=& \sum_{j\neq {\ell}}\frac{n_j{\bf
    e}_{\phi_{j}}}{|{\bf X}_{\ell}-{\bf
    X}_j|}+n_j\frac{\m}{2}\log|{\bf X}_{\ell}-{\bf
  X}_j|\left(\frac{d{\bf X}_j}{dT}\cdot 
  {\bf e}_{r_{j}}\right){\bf e}_{\phi_{j}}\nonumber\\
&&\mbox{ } -n_j\frac{\m}{2}(1+\log|{\bf X}_{\ell}-{\bf X}_j| 
)\left(\frac{d{\bf X}_j}{dT}\cdot {\bf e}_{\phi_{j}}\right){\bf 
  e}_{r_{j}}+\ordre(\m^2). 
\end{eqnarray*}
Note that the expression (\ref{eqLaw}) is accurate to all orders in
$\mu$; the expansion in $\mu$ is necessary only to evaluate $\del G$. 
To leading order in $\m$ the law of motion reads
\begin{equation}
\frac{d{\bf X}_{\ell}}{dT} \sim \frac{2}{n_{\ell}}\sum_{j\neq
  {\ell}}\frac{n_j{\bf e}_{r_{j}}}{|{\bf X}_{\ell}-{\bf X}_j|}. 
\label{lawq0}
\end{equation}

In particular, for two vortices at positions $(X_1,0)$ and $(X_2,0)$ with $X_1<X_{2}$
the laws of motion 
are given by
\beqa
\frac{dX_1}{dT}&=&\frac{n_2}{n_1}\left(\frac{2}{X_1-X_2},0\right)+\ordre(\m),\\
\label{2q=0}
\frac{dX_2}{dT}&=&\frac{n_1}{n_2}\left(\frac{2}{X_2-X_1},0\right)+\ordre(\m).
\label{2q=0a}
\eeqa
The direction of motion is always along the line of centres, with like
vortices repelling and opposites attracting.

\subsection{An alternative matching procedure}
\label{alternative}
The analysis above is based follows
the method that was used in \cite{neu90}. However, when $q>0$ there is
the 
added complication of three small
parameters, $\m$, $\e$ and $q$, 
rather than just two, $\m$ and $\e$. This has several implications on
the way we compute the asymptotic expansions, 
and affects the way the matching procedure must be carried out.

In the analysis above the key step which determines the law of motion
is matching the 
first-order inner solution   $\chi_1$ with the correction to the outer
phase due to the other vortices. When $q=0$ we are fortunate
that we do not need to expand the inner $\chi_1$ in powers of $\mu$,
so that we in effect retain all orders of $\mu$ while matching in
$\eps$. For $q>0$ we are not so lucky, and we need to expand both the
inner and outer 
solutions in powers of $\mu$ to make any progress. Since we then no
longer have the full $\mu$ expansion of either region, we cannot match
them together. We can get around this deficiency by writing down and
solving equations for the {\em outer limit} of the inner
solution. By solving these equations we are able to resum the infinite
series in $\mu$ that is present in the inner region, and this is
exactly what we need when matching. To illustrate this new procedure we
apply it here to the $q=0$ case, where we check the results against
the known solution above.

\subsubsection{Outer limit of the leading-order inner}
Rather than solving the inner equations and writing the solution in
terms of the outer variable and expanding, we rewriting the
leading-order inner {\em equations} in terms of 
the outer  
variable $R = \e r$ to obtain
\begin{eqnarray}
0 &=& \e^2(\n^2 f_0 -f_0|\n\ki_0|^2)+(1-f_0^2)f_0,\label{Jinout1}\\
0 &=& \e^2\n\cdot (f_0^2\n\ki_0). \label{Jinout2}
\end{eqnarray}
We now expand in powers of $\eps$ to give the outer limit of the
leading-order inner
solution as 
\beqas
\ki_0 &\sim& \widehat{\ki}_{00}
+\e^2\widehat{\ki}_{01}+\cdots\\
f_0 &\sim& \widehat{f}_{00} +\e^2\widehat{f}_{01}+\cdots. 
\eeqas
Substituting these expansions into (\ref{Jinout1}), (\ref{Jinout2}) gives
\begin{eqnarray}
\widehat{f}_{00} &=& 1,\qquad
\widehat{f}_{01} = -\frac{1}{2} |\n
\widehat{\ki}_{00}|^2,\\
\label{Jhatki0}
0 &=& \n^2\widehat{\ki}_{00},
\end{eqnarray}
with solution  $\widehat{\ki}_{00}= n_k\phi$.

\subsubsection{Outer limit of the first-order inner}
We  write the first-order inner equation in terms of the outer
variable to give
\begin{eqnarray}
- \eps \m \frac{d{\bf X}_k}{dT}\cdot\n f_0 &=&\eps^2 \n^2 f_1 
- \eps^2 f_1|\n  \ki_0|^2
-2 \eps^2 f_0 \n  \ki_0 \cdot \n \chi_1+
f_1 - 3 f_0^2 f_1,\\
- \m \eps f_0^2  \frac{d{\bf X}_k}{dT}\cdot\n \ki_0 &=& \eps^2 \n
 \cdot(f_0^2 \n
  \ki_1)+\eps^2\n \cdot(2 f_0 f_1\n
  \ki_0).
\end{eqnarray}
We now expand in powers of $\eps$ as 
\beqas
\ki_1 &\sim& \frac{\widehat{\ki}_{10}}{\eps} 
+\widehat{\ki}_{11}+\cdots,\\
f_1 &\sim& \widehat{f}_{10} +\e\widehat{f}_{11}+\cdots,
\eeqas
to give
\beqas
\widehat{f}_{10} & = & 0,\\
\widehat{f}_{11} & = & - \del \widehat{\ki}_{00} \cdot \del
\widehat{\ki}_{10},\\ 
    \n^2
  \widehat{\ki}_{10} & = & - \m  \frac{d{\bf X}_k}{dT}\cdot\n
  \widehat{\ki}_{00}= - n_k\m  \frac{d{\bf X}_k}{dT}\cdot\frac{1}{R} \,
  {\bf e}_{\phi}
= n_k \frac{\mu}{R} ( V_1 \sin \phi - V_2 \cos \phi),
\label{Jchi1a}
\eeqas
where we have written
\[
 \frac{d{\bf X}_\ell}{dT} = (V_{1},V_{2}).\]
Thus
\[ \widehat{\ki}_{10} = n_k\frac{\mu R \log R}{2} ( V_1 \sin \phi - V_2
\cos \phi) =  - n_k\frac{\mu R \log R}{2} \frac{d{\bf X}_k}{dT}\cdot 
  {\bf e}_{\phi}
\]
plus a homogeneous solution which comes from matching with  the outer.
We see that this homogeneous solution is
\[  {\bf X}\cdot \del G ({\bf X}_k),\]
giving
\beq
 \widehat{\ki}_{10} =  - n_k\frac{\mu R \log R}{2} \frac{d{\bf X}_k}{dT}\cdot 
  {\bf e}_{\phi} +{\bf X}\cdot \del G ({\bf X}_k).\label{moo1}
\eeq
This expression is the outer limit of the
 full $\mu$ expansion of the first-order inner solution.
Rewriting in terms of the inner variable gives
\[ \ki_{1} \sim  - n_k\frac{\mu r \log \eps }{2} \frac{d{\bf X}_k}{dT}\cdot 
  {\bf e}_{\phi} +{\bf x}\cdot \del G ({\bf X}_k) +\cdots,
\]
which is  (\ref{chi1}) as expected. Note that first
term in (\ref{moo1}), which is the particular integral, 
was obtained previously from the 
 $O(\mu)$ terms in the outer solution; this time we have used only the
leading-order outer solution. Thus this method allows us to match with
the inner solution when we only know a few terms in the
$\mu$-expansion of the outer solution.

\section{Equilibrium spiral wave solutions}
\label{sim}
We now consider equation \eqref{CGL} and  analyse equilibrium spiral
wave solutions in the limit where the parameter $q$ is small. These
equilibria, which correspond to single spirals with arbitrary winding
numbers, were  studied by Hagan \cite{hagan82} who showed
that the asymptotic wavenumber $k$, and thus the frequency of the
corresponding periodic solution to \eqref{CGL0}, is uniquely
determined by the parameter $q$. We shall recast Hagan's results in a
more systematic asymptotic framework in a way which generalises to the
many-spiral solutions considered in \S\ref{canonical} and \S\ref{middle}.

With $\p = f e^{i\chi}$ and $f$ and $\chi$ real,  we
seek solutions of the form $f=f(r)$   and
$\chi=n\phi+\varphi(r)$ 
with $f(r)$ and $\varphi_{r}(r)$ bounded as $r\to \infty$
and 
\[ \varphi_r(0) = 0, \qquad f(r) \sim C r^{n} \mbox{ as }r \to 0,\]
 where
$C$ is some positive constant, and a subscript denotes partial
differentiation. In fact it can be shown \cite{hagan82} 
that bounded solutions of \eqref{CGL}  satisfy $f(r) \to (1-k^2)^{1/2}$
and $\varphi_{r} \to -k$ as $r \to \infty$. 

 We start by introducing an auxiliary parameter $\eps$. For the
multi-spiral case $\eps$  will represent the inverse of the spiral
separation; here it represents the inverse of the outer core radius,
which will be defined shortly.
Rescaling \eqref{CGL} onto this new (outer) lengthscale by setting ${\bf X}
= \e {\bf x}$ gives
\begin{equation}
0 = \e^2 \n^2\p+(1+iq)(1-|\p|^2)\p-
\frac{i\e^2\alpha^2}{q}\p,
\label{outeralfaeq}
\end{equation}
where we have introduced the new parameter $\alpha = qk/\e$.
This can now be seen as an eigenvalue problem for $\alpha(q)$ which 
 provides the relationship between $k$ and $q$. The outer core radius
 of the spiral is the value of $\eps$ which makes $\alpha$ of order
 one as $q \ra 0$.

\subsection{Outer Region}
\label{simouter}

With $f=f(r)$   and
$\chi=n\phi+\varphi(r)$ 
equation (\ref{outeralfaeq}) becomes
\beqa
0 &=&  \e^2 \left(f'' + \frac{f'}{R}\right)
  -\e^2 f \left(\frac{n^2}{R^2} + (\varphi')^2\right)+
  (1-f^2)f,\label{outer1}\\  
0  &=&  \frac{\e^2}{R} (R f^2 \varphi')'
+qf^2(1-f^2)-\frac{\e^2\alpha^2}{q}f^2.\label{outer2}
\eeqa
Expanding in powers of $\eps$ as 
\beqas
f &\sim& f_0({\bf X}; q) + \e^2  f_1({\bf X}; q) +\cdots,\\
 \varphi &\sim& \varphi_0({\bf X}; q)+ \e^2 \varphi_1({\bf X};
 q) + \cdots, 
\eeqas
 we find
\[
 f_0 =  1,\quad f_1  = -\frac{1}{2}\left( \frac{n^2}{R^2} +
  (\varphi')^2\right),\]
\beq
\varphi_0''+
  \frac{\varphi_0'}{R}+
q\left(\frac{n^2}{R^2}+(\varphi_0')^2\right)-\frac{\alpha^2}{q}=0.  \label{phi0}
\eeq
Equation (\ref{phi0})  is  a Riccati  equation  and  can  be
  linearised  through  the 
transformation \linebreak\mbox{$\varphi_0 = (1/q)\log H_0$} to give
\begin{equation}
H_0''+\frac{H_0'}{R}+H_0\left(\frac{q^2n^2}{R^2}-\alpha^2\right)=0,
\end{equation}
with the general solution $H_0(R)=K_{inq}(\alpha R)+\lambda  I_{inq}(\alpha R)$
where $\lambda$ is an arbitrary real number, and $K_{inq}$ and $I_{inq}$ are the modified Bessel
functions of the first and second kind.  
Only when $\lambda=0$ is the function
$\varphi$ monotone \cite{hagan82}, so that the iso-phase contours are
spirals.
Then 
\begin{equation}
\ki_0\sim n\phi+\frac{1}{q}\log(K_{inq}(\alpha R)).
\label{ki0sim}
\end{equation}

\subsection{Inner region}
We return to the inner (original) scaling by setting  ${\bf X} =
\eps{\bf x}$ to give 
\begin{align*}
0   &=  \n^2  f   -f|\n  \ki|^2+(1-f^2)f,\\   0  &=   \n  \cdot(f^2\n
\ki)+q(1-f^2)f^2-\frac{\e^2 \alpha^2f^2}{q}.
\end{align*}
Expanding 
\beqas
f &\sim& f_0+\e f_1+\e^2 f_2+\cdots,\\
\varphi &\sim& \varphi_0+\e\varphi_1+\e^2\varphi_2+\cdots
\eeqas
the leading-order equations are
\beqa
\label{inner1aeq}
f_0''+\frac{f_0'}{r}-f_0\left(\frac{n^2}{r^2}+
(\varphi_0')^2\right)+(1-f_0^2)f_0 
  &=& 0,\\
\label{inner2aeq}
f_0\left(\varphi_0''+\frac{\varphi_0'}{r}\right)+2f_0'\varphi_0'
+q(1-f_0^2)f_0 
&= &0.
\eeqa
We now expand the leading-order solution in $\eps$ in powers of the
small parameter $q$ as
\beqa
f_0  &\sim&  f_{00}  +f_{02}q^2+  f_{04}q^4+\cdots,\label{f0expinq}\\  
\varphi_0  &\sim&
\frac{\varphi_{00}}{q}+\varphi_{02}q+\varphi_{04}q^3+\cdots.\label{phi0expinq}
\eeqa
Substituting    these    expansions    into   \eqref{inner1aeq}    and
\eqref{inner2aeq} and equating powers of $q$ gives
\beqa
\label{ki0eq}
\varphi_{00}&=&D_0,\\
0 &=&
f_{00}''+\frac{f_{00}'}{r}-n^2\frac{f_{00}}{r^2}+(1-f_{00}^2)f_{00},
\label{f0eq}\\
\label{ki2eq}
 \varphi_{02}'    &=&     -\frac{1}{r    f_{00}^2}\int_{0}^{r}    s
  f_{00}^2(1-f_{00}^2) \, ds,
\eeqa
with boundary conditions 
\begin{align}  
f_{00}(0)=& 0, \quad \lim_{r\to\infty} f_{00}(r) =  1, \label{f0bc}
\end{align}  
where  $D_0$  is a   real constant  to  be determined by matching.

\subsection{Asymptotic matching}
\subsubsection{Outer limit of the inner}
From expressions \eqref{f0eq}-\eqref{ki2eq} we find that as $r\to \infty$
\begin{equation}
\label{bc1eq}
 \varphi_{02}' \sim -n^2\,  \frac{\log r +
  c_n}{r} + \cdots,
\end{equation}
where $c_n$ is a constant given by
\[
c_n                                                                   =
\lim_{r\to\infty}\frac{1}{n^2}\left(\int_0^{r}f_{00}^2(s)\Big(1-f_{00}(s)^2\Big)s
\, ds -n^2 \log(r)\right).
\]
However, since we will see that $q$ is logarithmic in $\eps$, 
in order to match with the outer expansion we need the outer
limit of the  full expansion in $q$ of the leading-order inner
solution in $\eps$.  This is found to  be of the
form
\begin{align}
\label{innercapouter1eq}
f_0  &\sim 1-\frac{1}{r^2}\sum_{i=0}^{\infty}\alpha_i\{qn^2(\log(r)+c_n)\}^{2i}+\cdots,\\
\label{innercapouter2eq}
\varphi_0'         &\sim
-\frac{1}{r}\sum_{i=0}^{\infty}\beta_i\{qn^2(\log(r)+c_n)\}^{2i+1}+\cdots,
\end{align}
where $\alpha_i  >0$ and $\beta_i >0$ are  constants  independent
of $q$ and $n$. 
The necessity of taking all the terms in $q$ when matching can be seen,
since 
expansions \eqref{innercapouter1eq} and \eqref{innercapouter2eq} are
asymptotic provided  $q(\log(r)+c_n)\ll 1$;
with $q$ of order $1/\log (1/\eps)$ and $r$ of order $1/\eps$ all the
terms are the same order.

In principal, when matching logarithmic expansions, the full
logarithmic expansion 
in both the inner and outer regions needs to be found. However, when
these are finally compared term by term they must both be written in either the
inner or outer variable. Thus, in fact, it is enough to have the full
logarithmic expansion in one of the two regions only, providing the expansions
are written in terms of the other variable when comparing terms.

For the single spiral case we have the full expansion in $q$ of the
leading-order outer solution (\ref{ki0sim}), 
but only the first few terms in the
$q$-expansion of the leading-order inner solution
(\ref{ki0eq})-(\ref{ki2eq}).
Thus we can complete the matching by writing in outer solution
(\ref{ki0sim}) in terms of the inner variable, expanding in $\eps$ and
$q$, and comparing terms with (\ref{ki0eq})-(\ref{ki2eq}).
However, for the multiple-spiral case we will not be able to solve to
outer problem to all orders in $q$, so such an approach will not be available.

Fortunately there is a method which allows us to sum the $q$
expansion of the inner solution (\ref{ki0eq})-(\ref{ki2eq}), and this
method is still applicable in the multispiral case.
As in \S \ref{alternative},
rather than solving the leading-order inner equations, writing the
solution in terms of the outer variable, and then expanding again
ready to match, the trick is to write the {\em equation} for the
leading-order inner solution in terms of the outer variable, and only then
solve it. Hagan \cite{hagan82} thinks of  this as  a middle region
  expansion, but actually we are just writing down the equations
satisfied by the outer limit of the inner expansion.

Thus we begin   by    rewriting   the   leading-order   inner   equations
\eqref{inner1aeq} and  \eqref{inner2aeq} in terms of  the outer variable
$R = \e r$ to obtain
\beqa
0 &= & \e^2\left(f_0'' + \frac{f_0'}{R} -f_0\left(\frac{n^2}{R^2}+
(\varphi_0')^2\right)\right)+(1-f_0^2)f_0,\label{inout1eq}\\ 
0  &= & \frac{\e^2}{R} (R f_0^2 \varphi_0')'
+qf_0^2(1-f_0^2). \label{inout2eq}
\eeqa
Note that these are similar to, but not identical to, the equations in
the outer region (\ref{outer1})-(\ref{outer2}).
We now re-expand in powers of $\eps$ as 
\beqa
\varphi_0 &\sim & \widehat{\varphi}_{00}(q)
+\e^2\widehat{\varphi}_{01}(q)+\cdots,\label{inchiouteq}\\
f_0  &\sim & 
\widehat{f}_{00}(q) +\e^2\widehat{f}_{01}(q)+\cdots.\label{infouteq}
\eeqa
The leading-order term  in this expansion, $\widehat{\varphi}_{00}(q)$, is
just the first term (in $\eps$)  in the outer expansion of the
leading-order  inner solution, but now it includes the resummed
contribution from  all the  terms in
$q$. Substituting (\ref{inchiouteq})-(\ref{infouteq}) into
(\ref{inout1eq})-(\ref{inout2eq}) gives 
\[
\widehat{f}_{00}  =  1,\quad   \widehat{f}_{01}  =  -\frac{1}{2} \left(\frac{n^2}{R^2}+
(\widehat{\varphi}_{00}')^2\right)^2,
\]
\beq
\label{hatki0eq}
\widehat{\varphi}_{00}''+
  \frac{\widehat{\varphi}_{00}'}{R}+
q\left(\frac{n^2}{R^2}+(\widehat{\varphi}_{00}')^2\right)=0. 
\eeq
Equation \eqref{hatki0eq}  is  again  a  Riccati equation  which   can  be
linearised   with  the  change   of  variable   $\widehat{\varphi}_{00}  =
(1/q)\log\widehat{H}_0$ to give
\begin{equation}
\widehat{H}_0''+\frac{\widehat{H}_0'}{R}+\frac{q^2n^2}{R^2}\widehat{H}_0 = 0,
\end{equation} 
with solution 
\beq
\widehat{H}_0 = A(q)\eps^{-iqn} R^{iqn}+B(q)\eps^{iqn} R^{-iqn}, \label{hatH0}
\eeq
where  $A$ and  $B$ are  constants  that depend  on $q$ and the factors $\eps^{\pm  iqn}$ have been
included  to facilitate  their  determination by  comparison with  the
solution in the  inner variable. One relationship between  $A$ and
$B$ 
is determined by writing 
 $\widehat{\varphi}_{00}$ in terms of  $r$, expanding in powers of $q$, and
comparing with (\ref{innercapouter2eq}). A second is given by matching
with the outer solution.
Expanding  the constants in
powers of $q$ 
as 
\beqa
A(q) &\sim &  \frac{1}{q} A_{0}+A_{1}  +q A_{2} +\cdots,\\
B(q) &\sim &
\frac{1}{q} B_{0}+B_{1} +q B_{2} +\cdots,
\eeqa
writing $\widehat{H}_0$ in terms of $r$, and expanding for small
 $q$ we find 
\beqas
\widehat{H}_0  &  =& A(q)  e^{iqn\log  r} +  B(q)  e^{-iqn\log  r}\\ 
&\sim& \left(\frac{1}{q} A_{0}+A_{1} +q A_{2} +\cdots\right)\left(1+iqn\log r
- \frac{q^2n^2 \log^2 r}{2}+\cdots\right)\nonumber\\
&&\mbox{ }+\left(\frac{1}{q}  B_{0}+B_{1} +q B_{2}+\cdots\right)\left(1-iqn\log
r        -    \frac{q^2n^2 \log^2 r}{2}+\cdots\right)\\
&\sim&\frac{A_{0}+B_{0}}{q}+A_{1}+B_{1}+(A_{0}-B_{0})in\log
r\nonumber\\
&&\mbox{ }+q\left(A_{2}+B_{2}+(A_{1}-B_{1})in\log
r-n^2\frac{(A_{0}+B_{0})}{2}\log^2 r\right)+\cdots,
\eeqas
so that
\beqas
  \widehat{\varphi}_{00}'(r)
=\frac{\widehat{H}_0'(r)}{q\widehat{H}_0(r)}
&\sim& 
\frac{(A_{0}-B_{0})ni}{r(A_{0}+B_{0})}+
q\left(\frac{(A_{1}-B_{1})in}{(A_{0}+B_{0})r}-n^2\frac{\log  
r}{r}\right.\\
&&\left.\mbox{ }+n^2\frac{(A_{0}-B_{0})^2}{(A_{0}+B_{0})^2}
\frac{\log 
r}{r}-\frac{i(A_{0}-B_{0})(A_{1}+B_{1})}{(A_{0}+B_{0})^2}
\frac{n}{r}\right) + \cdots.
\eeqas
Comparing with (\ref{ki0eq}), and (\ref{bc1eq}) we see that
\begin{align}
\label{A0eq}
A_{0}-B_{0} =& 0,\\
\frac{(A_{1}-B_{1})}{A_{0}+B_{0}}i =& -n c_{n}. 
\label{A1eq}
\end{align}
The  remaining  equations  determining  $A$  and $B$  will  come  from
matching with the outer region.

\subsubsection{Inner limit of the outer}
Since we have now been
able to sum the $q$-expansion in the inner, we do not need the full
$q$-expansion of the outer, but only the leading-order term. Since we
have used the same transformation from $\phi$ to $H$ in the inner and
the outer, it is easiest to perform the matching in terms of $H$.
We find
\beq
 H_0(R)= K_{inq}(\alpha R)
\sim K_0(\alpha    R)    + O(q^2).\label{leadsingleouter}
 \eeq
Thus the inner limit of the outer is 
\beq   
H_0(R)\sim-\log\frac{R\alpha}{2}-\gamma + \ordre(q^2)
\label{hexpanded}
\eeq 
 where $\gamma$ is Euler's constant.
This should  match with $\widehat{H}_0(R)$ given by (\ref{hatH0}).
Exanding $\widehat{H}_0(R)$ in powers of $q$ gives
\beq
\widehat{H}_0 
\sim  \frac{A_{0}e^{-iqn\log
\e}+B_{0}e^{iqn\log    \e}}{q} +\cdots
\label{Heq}
\eeq
Comparing \eqref{Heq} and \eqref{hexpanded} and using \eqref{A0eq} 
 we find that $e^{iqn\log \e}+e^{-iqn\log \e} = \ordre(q)$,
so that
\begin{equation}
q|n\log\e|=   \frac{\pi}{2} +\nu q +\ordre(q^2). 
\label{qeqneq}
\end{equation}
Under this condition (\ref{Heq}) becomes
\[
\widehat{H}_0 
\sim -|n|(A_{0}+B_{0})\log R -(A_{0}+B_{0})\nu+i \,\mbox{sign}(n)  (A_{1}-B_{1}).
\]
Comparing  with \eqref{hexpanded} we see
\beqas
|n|(A_{0} + B_{0}) &=& 1,\\
i \,\mbox{sign}(n) (A_{1}-B_{1})-(A_{0}+B_{0})\nu  &=& -\log\alpha+\log 2-\gamma.
\eeqas
Eliminating $A$ and $B$ using \eqref{A0eq} and (\ref{A1eq}) gives, finally
\begin{equation}
 \alpha =2e^{c_n+\nu/|n|-\gamma},
\label{alfaSim}
\end{equation}
so that the eigenvalue $\alpha$ is now determined, and is indeed an
order-one constant. Recalling that
$\alpha=qk/\eps$ and using \eqref{qeqneq} 
we find that the asymptotic wavenumber is given by 
\beq 
\label{kq}
k(q)=\frac{2}{q}\exp\left(c_{n}-\gamma-\frac{\pi}{2q|n|}\right)(1+o(1))
\eeq 
in agreement with  Hagan \cite{hagan82}. The
corresponding frequency of the pattern  is given by 
\begin{equation}
\omega=q(1-\e^2\al^2)=q-4qe^{2c_{n}-2\gamma-\pi/q|n|}(1+o(1)).
\label{freq}
\end{equation}
We see from the expression for the phase in the  outer
region that the outer core radius corresponds to the radius at which
the iso-phase lines have become essentially azimuthal.
We will see that for the many-spiral case this corresponds to the
interaction between spirals becoming perpendicular to the line of
centres.
For separations less than the outer core radius the interaction is
algebraic, but for larger separations the interaction rapidly becomes
exponentially small. Thus the outer core radius forms an effective
``region of influence'' for each spiral.

\section{Interaction of spirals at the canonical separation}
\label{canonical}

We now want to combine the methods presented in \S\ref{q=0} and \S\ref{sim}
to consider the interaction of well-separated spirals when $0<q\ll 1$.
In each of \S\ref{q=0} and \S\ref{sim} we had two parameters to relate
to each other ($\eps$ and $\mu$, and $\eps$
and $q$ respectively); here we will have to consider the relative
sizes of all three parameters $\eps$, $\mu$ and $q$.
We saw in \S\ref{sim} that there is an outer core radius at which $H$
 switches from logarithmic growth to exponential decay.
Since the interaction between spirals occurs as a result of the
phase, it is clear that the interaction will crucially depend on the
relative sizes of the spiral separation and the outer core radius.

We start by considering the distinguished limit in which the spirals
are separated by distances of the same order as the outer core radius
as $q \ra 0$. Since
this outer core radius varies  exponentially with winding number,
we now make the assumption that all winding numbers are $\pm 1$, so
that all spirals have the same outer core radius.
 This means that we again have $\alpha=kq/\e$ of order
one, and a spiral separation of order $e^{\pi/2q}$. 

We will show that 
 the spirals interact, to leading order, in the direction
perpendicular to the line of centres, with a velocity 
of order $\e^2/|\log\e|$ (as in \S\ref{q=0}).
 We will find that for a pair of spirals, in contrast to fluid
 vortices, 
the  direction of motion of
 each depends only on its own winding number, and not on the winding
 number
 of the
 other spiral.

\subsection{Outer region}
\label{canouter}
Since we are considering spirals separated by distances large compared
to the inner core radius, the solution in the vicinity of each spiral
will be a small perturbation of the single spiral solution.
The whole pattern will have an  associated frequency or
asymptotic wavenumber. However, in the multi-spiral case,  as we shall
show, this frequency is no 
longer constant but  varies slowly as the spirals move, tending to the
single spiral limit as the separation tends to infinity.
  Thus we need to remember that 
when we use equation \eqref{CGL} the eigenvalue $k$ (or equivalently
$\al$) may depend on time.

As before we rescale time and space in \eqref{CGL} by setting ${\bf X}
= \e {\bf x}$, $T = \m\e^2 t$, to give 
\begin{equation}
\e^2\m\p_T = \e^2
\n^2\p+(1+iq)(1-|\p|^2)\p-\frac{i\e^2\alpha^2}{q}\p\label{outeral}.  
\end{equation}
Writing $\p = f e^{i \ki}$ as usual and separating real and imaginary
parts in (\ref{outeral}) gives 
\begin{align}
\m \e^2 f_T &= \e^2 \n^2 f -\e^2 f |\n \ki|^2+ (1-f^2)f,\\
\m \e^2 f^2 \ki_T &= \e^2\n\cdot (f^2\n\ki)+qf^2(1-f^2)-\e^2\frac{\alpha^2}{q}f^2.
\end{align}
Expanding in powers of $\eps$ as 
\beqas
f &\sim& f_0({\bf X}, T; q,\m) + \e^2 f_1({\bf X}, T;q,\m) + \cdots,\\
\ki &\sim & \ki_0({\bf X}, T;q,\m) + \e^2 \ki_1({\bf X}, T;q,\m) +
\cdots,
\eeqas
we find
\begin{align}
 f_0 &= 1,\quad f_1 = -\frac{1}{2} |\n \ki_0|^2,\nonumber\\
 \m \ki_{0T} &= \n^2 \ki_0 + q|\n \ki_0|^2 - \frac{\alpha^2}{q}.\label{chieqn}
\end{align}
It is tempting now to linearise (\ref{chieqn}) via the Cole-Hopf
transformation $\chi_0 = (1/q) \log h$ as in \S\ref{sim}, to give
\begin{equation}
\label{h-eqn}
\mu h_T = \n^2 h - \alpha^2 h. 
\end{equation}
Then, since the equation is linear, we could sum up the contributions
from each spiral to give 
\begin{equation}
\label{Kiqn}
h = \sum_{j=1}^N \beta_{j}(T)e^{qn\phi_j}K_{iqn}(\alpha R_j),
\end{equation}
where $R_j$ and $\phi_j$ are the polar variables centred 
on the $j$\,th spiral and the weights $\beta_j$ depend on the 
slow time variable $T$. 
This function has the
right type of singularities to match with the spiral core when we
expand it locally. Unfortunately, when the transormation is undone to
return to $\chi_0$ and $\psi$, we see that $\psi$ is no longer single-valued.
This problem with multivaluedness is pointed out in  \cite{pismen03}
and \cite{pismen92}, where the authors rightly claim that it
 invalidates the use of the Cole-Hopf 
transformation. 
Nevertheless, in what follows we will show that this transformation
can  be used to advantage without causing  $\p$ to
become multivalued, providing care is taken. 
The key is the observation that for a single
spiral the dependence of $\chi$ on $\phi$ occurs at $O(1)$, not
$O(1/q)$, so that at leading order the Cole-Hopf transformation can be
used without difficulty. Then, at first order, the 
single-valueness of  $\p$ will be maintained by introducing exactly 
 the right multivalueness in $h$.

To simplify the exposition 
we now make the assumption, which will be justified {\em a
  posteriori}, that $\mu$ and $q$ are the same order as $\eps \ra 0$
 (we will see that both are $O(1/\log(\eps))$.  We therefore write
  $\mu = q \tilde{\mu}$ and treat $\tilde{\mu}$ as $O(1)$.
Expanding $\ki_0$ in powers of $q$ as 
\[
\ki_0
\sim \frac{\ki_{00}}{q}+\ki_{01}+\cdots\]
gives, to leading order,  
\begin{equation}
0= \n^2 \ki_{00} + |\n \ki_{00}|^2 - \alpha^2.
\label{ki00}
\end{equation}
Linearising  \eqref{ki00}  through the Cole-Hopf
transformation $\ki_{00} = \log h_0$ gives 
\begin{equation}
\label{h00}
0 = \n^2 h_0 - \alpha^2 h_0. 
\end{equation}
Now, at leading order, the solution for a single spiral is given by
(\ref{leadsingleouter}); for multiple spirals we can some these to give
\beq
h_0 = \sum_{j=1}^{N} \beta_{j}(T) K_0(\alpha R_j),\qquad 
\ki_{00}=\log \sum_{j=1}^{N} \beta_{j}(T) K_0(\alpha R_j).
\label{outer0}
\eeq
Note that, because the leading-order solution  does not depend on
$\phi$, there is no problem with multivalueness of $\p$. 
The weights
$\beta_{j}$ will be 
determined by matching with the inner expansion.

\subsection{Inner Region}
\label{caninner}
We rescale near the $\ell$\,th vortex by setting ${\bf X} = {\bf X}_{\ell} +
\eps {\bf x}$ to give
\begin{eqnarray}
\e\m \left(\e f_T - \frac{d{\bf X}_{\ell}}{dT}\cdot\n f\right) &=& \n^2 f -f|\n
  \ki|^2+(1-f^2)f\\
\e \m f^2\left(\e \ki_T-\frac{d{\bf X}_{\ell}}{dT}\cdot\n \ki\right) &=& \n \cdot(f^2\n
  \ki)+q(1-f^2)f^2-\frac{\e^2 \alpha^2f^2}{q}
\end{eqnarray}
or equivalently
\begin{equation}
\e\m\left(\e\p_T-\frac{d{\bf X}_{\ell}}{dT}\cdot\n\p\right) =
\n^2\p+(1+iq)(1-|\p|^2)\p-
i \frac{\e^2\alpha^2}{q}\p
\end{equation}
Expanding 
\beqas
f &\sim& f_0({\bf x};q,\m)+\e f_1({\bf x};q,\m)+\cdots,\\
\ki &\sim& \ki_0({\bf x};q,\m)+\e\ki_1({\bf x};q,\m)+\cdots,
\eeqas 
the leading-order equations are
\begin{eqnarray}
\label{inner1}
0 &=& \n^2 f_0 -f_0|\n\ki_0|^2+(1-f_0^2)f_0,\\
\label{inner2}
0 &=& \n\cdot (f_0^2\n\ki_0)+q(1-f_0^2)f_0^2,
\end{eqnarray}
or equivalently
\begin{equation}
\label{leadingInner}
0 = \n^2 \p_0 + (1+iq)\p_0(1-|\p_0|^2).
\end{equation}
The effect of the other spirals is felt at higher order, so that 
the leading-order inner solution is of the form
$f_0 = f_0(r)$ and $\ki_0 = n_{\ell}\phi + \varphi_0(r)$, where
\begin{eqnarray}
\label{inner1a}
f_0''+\frac{1}{r}f_0'-f_0\left(\frac{1}{r^2}+(\varphi_0')^2\right)+(1-f_0^2)f_0   &=& 0,\\
\label{inner2a}
f_0
\left(\varphi_0''+\frac{\varphi_0'}{r}\right)
+2f_0'\varphi_0'+q(1-f_0^2)f_0 &=& 0,
\end{eqnarray}
where are of course the equations for a single spiral
(\ref{inner1aeq})-(\ref{inner2aeq}). 
Expanding in powers of $q$ as in (\ref{f0expinq})-(\ref{phi0expinq})
we find that  
$\varphi_{00} = \varphi_{00}(T)$,
with $f_{00}$ and $\varphi_{02}$ satisfying (\ref{f0eq})-(\ref{f0bc}).

At first order in $\eps$ we find
\begin{equation}
\label{1orderinner}
-\m \frac{d{\bf X}_{\ell}}{dT} \cdot \n \p_0 = \n^2
 \p_1+(1+iq)(\p_1(1-|\p_0|^2)-\p_0(\p_0\p_1^*+\p_0^*\p_1)) 
\end{equation}
or equivalently
\begin{eqnarray}
- \m \frac{d{\bf X}_{\ell}}{dT}\cdot\n f_0 &=& \n^2 f_1 
-f_1|\n  \ki_0|^2
-2 f_0 \n  \ki_0 \cdot \n \chi_1+
f_1 - 3 f_0^2 f_1,\label{firstin1}\\
- \m f_0^2 \frac{d{\bf X}_{\ell}}{dT}\cdot\n \ki_0 &=& \n \cdot(f_0^2\n
  \ki_1)+\n \cdot(2 f_0 f_1\n
  \ki_0)
+2 q f_0 f_1 -4 q f_0^3 f_1.\label{firstin2}
\end{eqnarray}

The solvability condition on (\ref{1orderinner}) will give the law of
motion. We derive this condition in \S\ref{cansolv}. First, we match
the leading-order solutions in each region to determine the weights
$\beta_j$ and the eigenvalue $\alpha$.

\subsection{Asymptotic matching}
\subsubsection{Outer limit of the leading-order  inner}

We  use the same trick as in Section \ref{sim} to sum the 
$q$-expansion of the outer limit of the leading-order inner
solution. We begin by 
rewriting the leading-order inner equations \eqref{inner1} 
and \eqref{inner2} in terms of the outer variable 
$R_{\ell} = \e r$ to obtain
\begin{eqnarray}
0 &=& \e^2(\n^2 f_0 -f_0|\n\ki_0|^2)+(1-f_0^2)f_0,\label{inout1}\\
0 &=& \e^2\n\cdot (f_0^2\n\ki_0)+q(1-f_0^2)f_0^2. \label{inout2}
\end{eqnarray}
We will now denote $R_{\ell}$ by $R$ to simplify the notation.
We now expand in powers of $\eps$ as 
\beqas
\ki_0 &\sim& \widehat{\ki}_{00}(q)
+\e^2\widehat{\ki}_{01}(q)+\cdots,\\
f_0 &\sim& \widehat{f}_{00}(q) +\e^2\widehat{f}_{01}(q)+\cdots.
\eeqas
The leading order term in this expansion $\widehat{\ki}_{00}(q)$ is just
the first term (in $\eps)$ in the outer expansion of the leading order
inner solution, including all the terms in $q$. Substituting
these expansions into (\ref{inout1}), (\ref{inout2}) gives
\begin{eqnarray}
\widehat{f}_{00} &=& 1,\quad \widehat{f}_{01} = -\frac{1}{2} |\n
\widehat{\ki}_{00}|^2,\nonumber\\
\label{hatki0}
0 &=& \n^2\widehat{\ki}_{00} + q|\n\widehat{\ki}_{00}|^2. 
\end{eqnarray}
Equation \eqref{hatki0}  can be linearised 
with the usual change of variable $\widehat{\ki}_{00} = (1/q)\log\widehat{h}_0$
to give 
\[ \n^2\widehat{h}_0= 0.\]
As in the single spiral case the relevant solution is of the form 
$\widehat{\ki}_{00}= n_{\ell}\phi + \widehat{\varphi}(R)$ so that
$\widehat{h}_0=
e^{qn_{\ell}\phi}e^{q\widehat{\varphi}(R)} = e^{qn_{\ell}\phi}H_0(R)$ where
\begin{equation}
H_0''+\frac{H_0'}{R}+q^2\frac{H_0}{R^2} = 0,
\end{equation} 
with solution
\begin{equation}
\label{solutionH}
H_0 = A_{\ell}(q,T)\eps^{-iq n_{\ell}} R^{iq n_{\ell}}+B_{\ell}(q,T)\eps^{iq n_{\ell}} R^{-iq n_{\ell}},
\end{equation}
where the constants $A_{\ell}$ and $B_{\ell}$ may depend not only on $q$
but also on the slow time $T$, and  may be
different at each vortex; as before the factors $\eps^{\pm iq
  n_{\ell}}$ have been 
included to facilitate 
comparison with the solution in the inner variable. 
Expanding
\beqa
A_{\ell}(q)&\sim& \frac{A_{\ell 0}}{q} +A_{\ell 1} +q A_{\ell 2} +\cdots,\label{Al}\\
B_{\ell}(q)& \sim& \frac{B_{\ell 0}}{q} +B_{\ell 1} +q B_{\ell 2}
+\cdots, \label{Bl}
\eeqa  
writing $\hat{\chi}_{00}$ in terms of $r$, expanding in powers of $q$, and
comparing
with (\ref{ki0eq}), and (\ref{bc1eq})
as in the single spiral case gives
\begin{align}
\label{A0}
A_{\ell 0}-B_{\ell 0} =& 0,\\
\label{A1}
\frac{(A_{\ell 1}-B_{\ell 1})}{A_{\ell 0}+B_{\ell 0}}i =& -n_{\ell}c_{n_{\ell}} \quad \textrm{for $\ell =
  1, \ldots, N$}. 
\end{align}
The remaining equations determining $A_{\ell}$ and $B_{\ell}$ will be fixed when
matching with the outer region.

\subsubsection{Inner limit of the outer}
 To compute the inner limit of the
leading-order outer solution we rewrite solution \eqref{outer0} in
terms of the inner variable by setting 
${\bf X} = {\bf X}_{\ell}+\e {\bf x}$ and expand in powers of $\e$ to give
\beqa
h_0  &=& \sum_{j=1}^{N}\beta_{j}(T) K_0(\alpha |{\bf X}_{\ell}+\e
{\bf x}-{\bf X}_j|)\nonumber\\
&\sim &-\beta_{\ell} \log\frac{\alpha \eps
  r}{2}-\beta_{\ell}\gamma+G({\bf X}_{\ell})+\e {\bf
  x}\cdot\nabla G({\bf X}_{\ell})+\cdots,\label{h0} 
\eeqa
where 
\beq
G({\bf X}) = \sum_{j=1,j\neq\ell}^{N} \beta_{j}(T) K_0(\alpha |{\bf
  X}-{\bf X}_j|).\label{G0eqn}
\eeq

\subsubsection{Leading order matching: determination of the frequency}
We now match the inner limit of the leading-order outer solution (\ref{h0}) with
the outer limit of the leading-order inner solution \eqref{solutionH}.
Since we have summed the $q$-expansion 
of the outer limit of the leading-order inner, while we have only the
first term in the $q$-expansion of the inner limit of the
leading-order outer, we must write both expansions in the outer
variable $R$ before comparing terms.

The   $q$-expansion of the  inner limit of
the leading-order outer solution is 
\beq
\ki_0  \sim  \frac{\log h_0}{q}+\cdots 
\sim \frac{1}{q}\log\left(-\beta_{\ell} \log\frac{\alpha R}{2}-\beta_{\ell}\gamma+G({\bf X}_{\ell})\right)+\cdots 
\label{inout}
\eeq
The  outer limit of the leading-order inner solution is
\beqa
\widehat{\ki}_{00}&=& n_{\ell}\phi + \frac{1}{q}\log H_0(R) \nonumber\\
&= &n_{\ell}\phi+ 
\frac{1}{q}\log \left(A_{\ell}(q)\eps^{-iq n_{\ell}} R^{iq
  n_{\ell}}+B_{\ell}(q)\eps^{iq n_{\ell}} R^{-iq n_{\ell}}\right) \nonumber\\
&= &n_{\ell}\phi+ 
\frac{1}{q}\log \left(\frac{A_{\ell 0}e^{-iqn_{\ell}\log \e}+B_{\ell 0}e^{iqn_{\ell}\log \e}}{q}\right)+\cdots. 
\label{outinone}
\eeqa
Comparing (\ref{outinone}) and (\ref{inout}) we see that, as in the
single spiral case, we require
\begin{equation}
e^{iqn_{\ell}\log \e}+e^{-iqn_{\ell}\log \e} = \ordre(q),
\end{equation}
so that, since $n_{\ell} = \pm 1$,  
\begin{equation}
q\log1/\e= \frac{\pi}{2}+\nu q+\ordre(q^2).
\label{qeqn}
\end{equation}
This is the relationship between $q$ and $\eps$ required  for $\alpha$ to be of
order one, and is equivalent to assuming that the typical spiral
separation  $1/\e=\ordre(e^{\pi/2q})$.

Assuming (\ref{qeqn}) holds, equation (\ref{outinone}) becomes
\begin{equation} 
\widehat{\ki}_{00}\sim  \frac{1}{q}\log\left(-(A_{\ell
    0}+B_{\ell 0})\log R -(A_{\ell 0}+B_{\ell 0})\nu+i
  \,\mbox{sign}(n_{\ell})  (A_{\ell 1}-B_{\ell 1})\right)+\cdots .  
\label{outin}
\end{equation} 
Comparing  with (\ref{inout}) 
gives 
\beqas
A_{\ell 0}+B_{\ell 0}&=&\beta_{\ell},\\
i \,\mbox{sign}(n_{\ell}) (A_{\ell 1}-B_{\ell 1})-(A_{\ell 0}+B_{\ell
  0})\nu  &=& 
-\beta_{\ell}(\log\alpha-\log 2+\gamma) +G({\bf X}_{\ell}). 
\eeqas
Eliminating $A_{\ell}$ and $B_{\ell}$ using \eqref{A0} and \eqref{A1}
and \eqref{G0eqn}  gives
\begin{equation}
-(c_{1}+\nu)\beta_{\ell} = -\beta_{\ell}(\log\alpha-\log 2+\gamma) + 
\sum_{j\neq\ell}^{N}\beta_{j} K_0(\alpha |{\bf X}_{\ell}-{\bf X}_j|).\label{90}
\end{equation}
Since (\ref{90}) holds for each spiral this  is a system of $N$
linear  equations for the unknown weights 
 $\beta_{j}$, $j = 1,\ldots,n$. Since the system of equations is
 homogeneous, a non-zero solution will exist only if the determinant
 is zero: this is the condition which determines the eigenvalue $\al$
 in the multispiral case. Note that the weights $\beta_{j}$ and the
 eigenvalue $\al$ (and therefore $k$) depend on the position of the
 spiral centres, and will therefore evolve on the slow timescale $T$.

For the case of just  two spirals equations (\ref{90}) become
\begin{align}
-(c_{1}+\nu)\beta_{1} =& -\beta_{1}(\log\alpha-\log 2+\gamma) + \beta_{2}
K_0(\alpha |{\bf X}_1-{\bf X}_2|),\\ 
-(c_{1}+\nu)\beta_{2} =& -\beta_{2}(\log\alpha-\log 2+\gamma) + \beta_{1}
K_0(\alpha |{\bf X}_1-{\bf X}_2|), 
\end{align}
from which we see that $\beta_{1} = \beta_{2}$ and the eigenvalue condition
is just 
\begin{equation}
-c_{1}-\nu= -(\log\alpha-\log 2+\gamma) + K_0(\alpha |{\bf X}_1-{\bf X}_2|),
\label{alfa2s}
\end{equation}
so that
\beq
\alpha=2 e^{c_1+ \nu-\gamma+K_0(\alpha |{\bf X}_1-{\bf X}_2|)}, \qquad
k = \frac{2}{q} \,e^{-\pi/2q+c_1-\gamma+ K_0(\alpha |{\bf X}_1-{\bf
    X}_2|)}. \label{alphaK}
\eeq
This expression for $k$ agrees with that derived for a pair of spirals
in \cite{pismen92} by other
methods.

We observe that as the spiral separation $|{\bf X}_1-{\bf
  X}_2|\ra\infty$, 
$\alpha$ approaches that corresponding to a single spiral of unitary
winding number \eqref{alfaSim}. 

Thus far we have matched at leading order and determined the
eigenvalue $\alpha$ and corresponding asymptotic wavenumber $k$. To
determine a law of motion of the individual spirals we need to match
at first order.

\subsubsection{Outer limit of the first-order inner}
\label{4.3.4}
We sum the $q$-expansion of the outer limit of the  first-order inner
solution in the same way that we did for the outer limit of the
leading-order inner solution.
We first write equation (\ref{firstin1})-(\ref{firstin2}) in terms of the outer
variable to give
\begin{align*}
- \eps \m \frac{d{\bf X}_{\ell}}{dT}\cdot\n f_0 &=\eps^2 \n^2 f_1 
- \eps^2 f_1|\n  \ki_0|^2
-2 \eps^2 f_0 \n  \ki_0 \cdot \n \chi_1+
f_1 - 3 f_0^2 f_1,\\
- \m \eps f_0^2  \frac{d{\bf X}_{\ell}}{dT}\cdot\n \ki_0 &=\eps^2 \n
 \cdot(f_0^2 \n
  \ki_1)+\eps^2\n \cdot(2 f_0 f_1\n
  \ki_0)
+2 q f_0 f_1 -4 q f_0^3 f_1.
\end{align*}
We now expand in powers of $\eps$ as 
\beqas
\ki_1 &\sim& \frac{\widehat{\ki}_{10}(q,\mu)}{\eps} 
+\widehat{\ki}_{11}(q,\mu)+\cdots,\\
f_1 &\sim& \widehat{f}_{10}(q,\mu) +\e\widehat{f}_{11}(q,\mu)+\cdots,
\eeqas
to give
\begin{align}
\widehat{f}_{10} & =  0,\quad\widehat{f}_{11}  =  - \del\widehat{\ki}_{00} \cdot \del
\widehat{\ki}_{10},\nonumber\\
- \m  \frac{d{\bf X}_{\ell}}{dT}\cdot\n \widehat{\ki}_{00} &=  
  \n^2  \widehat{\ki}_{10} -2 q  \widehat{f}_{11} = 
\n^2 \widehat{\ki}_{10}+ 2 q \del \widehat{\ki}_{00} \cdot \del
\widehat{\ki}_{10}.\label{chi1a}
\end{align}
Motivated by the transformation we applied to $\widehat{\ki}_{00}$ we
write
\[ \widehat{\ki}_{10} = \frac{\widehat{h}_1}{q \widehat{h}_0} = \frac{\widehat{h}_1 e^{- q
    \widehat{\ki}_{00}}}{q }, 
\]
whence (\ref{chi1a}) becomes 
\beq
- \m  \frac{d{\bf X}_{\ell}}{dT}\cdot\n \widehat{h}_0 =
\del^2 \widehat{h}_1,
\label{eqki1}
\eeq
where we have written $\widehat{\ki}_{00}$ in terms of $\widehat{h}_0$.
Recalling that $\widehat{h}_0 = e^{qn_{\ell}\phi} H_0(R)$, and
denoting as before
\[ \frac{d{\bf X}_\ell}{dT} = (V_{1\ell},V_{2\ell}),\]
 the left-hand
side of \eqref{eqki1} gives
\beqas
\lefteqn{ - \m  \frac{d{\bf X}_{\ell}}{dT}\cdot \left(\frac{qn_{\ell} e^{qn_{\ell}\phi} H_0(R)}{R}
{\bf e}_{\phi} + H_0'(R) e^{qn_{\ell}\phi}{\bf e}_{R}\right)}&&\\
& = & \m  e^{qn_{\ell}\phi} \sin \phi   
 \left( qn_{\ell} V_{1\ell} \frac{H_0(R)}{R} - V_{2\ell}
 H_0'(R)\right)
-\m e^{qn_{\ell}\phi}\cos \phi \left(qn_{\ell}  V_{2\ell} \frac{H_0(R)}{R} + V_{1\ell}
 H_0'(R)\right)\\
&=&\m  e^{qn_{\ell}\phi} q n_{\ell}\sin\phi\big(R^{iqn_{\ell} -1} A_{\ell}\eps^{-iqn_{\ell}} (V_{1\ell}   - i  V_{2\ell})+ R^{-iqn_{\ell} -1}B_{\ell} \eps^{iqn_{\ell}}
( V_{1\ell}  + i V_{2\ell})\big)
\\&&\mbox{ }
-\m e^{qn_{\ell}\phi}q n_{\ell}\cos \phi\big(R^{iqn_{\ell} -1} A_{\ell}\eps^{-iqn_{\ell}}( V_{2\ell}   + i  V_{1\ell})
-R^{-iqn_{\ell} -1}B_{\ell} \eps^{iqn_{\ell}}  ( V_{2\ell}  - i V_{1\ell}))\\
&=& \m  e^{qn_{\ell}\phi}\frac{(e^{i \phi} - e^{- i \phi})}{2i} R^{iqn_{\ell} -1}qn_{\ell} A_{\ell}\eps^{-iqn_{\ell}}
 \left(  V_{1\ell}   - i  V_{2\ell}  \right)\\
&& \mbox{ }+\m  e^{qn_{\ell}\phi} \frac{(e^{i \phi} - e^{- i \phi})}{2i} R^{-iqn_{\ell} -1} qn_{\ell} B_{\ell} \eps^{iqn_{\ell}}
\left( V_{1\ell}  + i V_{2\ell}   \right)\\
&& \mbox{ }-\m e^{qn_{\ell}\phi}\frac{(e^{i \phi} + e^{- i \phi})}{2}R^{iqn_{\ell} -1} qn_{\ell} A_{\ell}\eps^{-iqn_{\ell}}\left( V_{2\ell}   + i  V_{1\ell}  \right)\\
&& \mbox{ }-\m e^{qn_{\ell}\phi}\frac{(e^{i \phi} + e^{- i \phi})}{2}   R^{-iqn_{\ell} -1}qn_{\ell} B_{\ell} \eps^{iqn_{\ell}}  \left( V_{2\ell}  - i V_{1\ell}
\right) 
\\ &=&\mbox{ }-\frac{\m q n_{\ell} e^{qn_{\ell}\phi}}{R}\big(e^{i \phi} R^{iqn_{\ell}} A_{\ell}\eps^{-iqn_{\ell}}(V_{2\ell}+i V_{1\ell})
-e^{- i \phi} R^{-iqn_{\ell}}B_{\ell} \eps^{iqn_{\ell}} ( V_{2\ell}  - i V_{1\ell})\big). 
\eeqas
Therefore, writing 
\[
\widehat{h}_1 = - \mu qn_{\ell} A_{\ell} \eps^{-i
  qn_{\ell}}(V_{2\ell}+i V_{1\ell}) \g_1(R)\, e^{(qn_{\ell}+i) \phi}  
- \mu qn_{\ell} B_{\ell} \eps^{i qn_{\ell}}(V_{2\ell}-i V_{1\ell}) \g_2(R)\, e^{(qn_{\ell}-i) \phi} , \]
gives
\beqas
 \g_1'' + \frac{\g_1'}{R} + \frac{(qn_{\ell}+i)^2\g_1}{R^2} &=&  R^{
  iqn_{\ell} -1},\\ 
 \g_2'' + \frac{\g_2'}{R} + \frac{(qn_{\ell}-i)^2\g_2}{R^2} &=&  R^{
  -iqn_{\ell} -1}, 
\eeqas 
with the general solution
\begin{align*}
 \g_1 &= \frac{R^{ iqn_{\ell} +1}}{4 i
  qn_{\ell}}+\gamma_1R^{1-iqn_{\ell}}+
\gamma_3 R^{-1+iqn_{\ell}},\\ 
 \g_2 &= -\frac{R^{ -iqn_{\ell} +1}}{4 i
  qn_{\ell}}+\gamma_2R^{1+iqn_{\ell}}+
\gamma_4R^{-1-iqn_{\ell}}, 
\end{align*}
where $\gamma_1$, $\gamma_2$, $\gamma_3$ and $\gamma_4$ are arbitrary
constants. By comparison with the inner solution we find that
$\gamma_3=\gamma_4=0$; $\gamma_1$ and
$\gamma_2$,  will be determined by matching to the inner limit of
the outer solution. 
Thus  (with a redefinition of $\gamma_1$ and $\gamma_2$) the outer limit
of the first-order inner solution 
is given by
\beqa
\widehat{h}_1 &=& \mbox{}  -\frac{\mu  A_{\ell} \eps^{-i
    qn_{\ell}}(V_{1\ell}-i V_{2\ell})}{4  }
R^{ iqn_{\ell}
  +1}\, e^{(qn_{\ell}+i)\phi}
- \frac{\mu B_{\ell} \eps^{i qn_{\ell}}( V_{1\ell}+ i V_{2\ell})}{4  } R^{ -iqn_{\ell} +1}\, 
e^{(qn_{\ell}-i)\phi}
\nonumber \\  
&&\mbox{ }
+\gamma_1R^{1-iqn_{\ell}}\, e
^{(qn_{\ell}+i)\phi}
+\gamma_2R^{1+iqn_{\ell}}\, 
e^{(qn_{\ell}-i)\phi}.\label{outinh1}
\eeqa

\subsubsection{First order matching: determination of the law of motion}

To determine the law of motion we need to match the two-term inner
expansion to the one-term outer expansion (in the notation of Van
Dyke \cite{vandyke}, we impose (2ti)(1to)=(1to)(2ti)).
Since we have summed the $q$ expansion of the inner, but have only the
first term in the $q$ expansion of the outer, we again need to compare
both series in terms of the outer variable.

The outer limit of the two-term expansion in the inner region is
\[ \chi \sim \frac{1}{q}\log \hat{h}_0 + \frac{\eps \hat{h}_1}{q
  \hat{h}_0} \sim \frac{1}{q}\log \left(\hat{h}_0 + \eps
  \hat{h}_1\right).
\]
Thus, at leading order in $q$, we can perform the matching using $h$
as in \S\ref{sim}. Comparing (\ref{outinh1}) with (\ref{h0}) we see
that we need $\gamma_1$ and $\gamma_2$ to be of order one, in which
case (with the usual expansion $\gamma_i \sim \gamma_{i0} + q
\gamma_{i1} + \cdots$ for $\gamma$ and $V$), and recalling (\ref{Al})-(\ref{Bl})
\beqas
\widehat{h}_1 &\sim& \mbox{}  \left(\gamma_{10}-\frac{\tilde{\mu}
    A_{\ell0}
 e^{in_{\ell}\pi/2}(V_{1\ell0}-i V_{2\ell0})}{4  }\right)
R \, e^{i\phi}
+\left(\gamma_{20}- \frac{\tilde{\mu} B_{0\ell}  e^{-in_{\ell}\pi/2}
( V_{1\ell0}+ i V_{2\ell0})}{4  } \right)R\, 
e^{-i\phi}.
\eeqas
and in order to match with (\ref{h0}) we require this to be equal to
$ {\bf
  X}\cdot\nabla G({\bf X}_{\ell})$ so that
\beqas
 \gamma_{10} &=& \frac{\tilde{\mu}  A_{\ell0}  e^{in_{\ell}\pi/2}
(V_{1\ell0}-i V_{2\ell0})}{4  }+\frac{G_{X}({\bf X}_{\ell}) - i G_{Y}({\bf
    X}_{\ell})}{2}, \\
\gamma_{20} &=& \frac{\tilde{\mu} B_{0\ell}  e^{-in_{\ell}\pi/2}
( V_{1\ell0}+ i V_{2\ell0})}{4  } +\frac{G_{X}({\bf X}_{\ell}) + i G_{Y}({\bf
    X}_{\ell})}{2}.
\eeqas
Now, writing (\ref{outinh1}) in terms of the inner variable $r$ and
expanding in powers of $q$ we find that, as $r \ra \infty$,
\beq
 \ki_{10} 
 \sim 
-\frac{\tilde{\mu} r}{2}\,(V_{1\ell0} \cos \phi +V_{2\ell0}
\sin\phi)+\frac{n_{\ell} r}{\beta_{\ell}}\n G ({\bf X}_{\ell})\cdot
	{\bf e}_{\phi}. 
\label{firstinner}
\end{equation}

\subsection{Solvability Condition}
\label{cansolv}
We recall that equation \eqref{1orderinner} is a linear equation of the form
\[
L(q,\p_0)[\p_1] = b,
\] 
where
\beqa
\label{operator}
L(q,\p_0)[\p_1] &=& (1-iq)\n^2 \p_1+\p_1(1-|\p_0|^2)-\p_0(\p_0\p_1^*+\p_0^*\p_1),\\
b& =& -\m \frac{d{\bf X}_{\ell}}{dT} \cdot \n \p_0.
\eeqa
In the case of nonzero $q$ the operator $L$ is no longer self adjoint.
However, the adjoint operator is given by
\begin{equation}
\label{adjoint}
L^*(q,\p_0)[v] = (1+iq)\n^2 v+v(1-|\p_0|^2)-\p_0(\p_0 v^*+\p_0^* v) = 
L(-q,\p_0)[v] .
\end{equation}
Choosing $v$ to be a non-trivial solution to the homogeneous equation
$L(-q,\p_0)[v] = 0$ and  using the Fredholm Alternative (integrating
by parts) we obtain
the solvability condition 
\begin{equation}
-\int_D \Re\left\{\m\frac{d{\bf X}_{\ell}}{dT}\cdot \n \p_0
  v^*\right\} dD=\int_{\partial 
  D}\Re\left\{(1-iq)\left(v^*\frac{\partial \p_1}{\partial
    n}-\frac{\partial v^*}{\partial n}\p_1\right)\right\}\, dl,
\label{solvability}
\end{equation}
where $D$ is a ball of radius $r$.
The non-trivial solution of the adjoint equation
are directional derivatives
 $\n\p_0 \cdot {\bf d}$ of $\p_0$, with $q$ replaced by $-q$, where
${\bf d}$ is any vector in $\mathbb{R}^2$. 
Since we are only interested in the leading-order law of motion, we
expand the first-order inner equation (\ref{operator}) in powers of
$q$, and consider the solvability condition on the first term in this
expansion. The leading-order (in $q$) operator is self-adjoint, and
the non-trivial solutions are derivatives of the leading-order
($\eps=0$, $q=0$) inner problem. At leading order in $q$ the
solvability condition is exactly that of \S\ref{q=0solv}, namely 
\begin{equation}
\int_0^{2 \pi}({\bf e}_{\phi} \cdot {\bf d}) \left( 
  \frac{\dd \chi_{10}}{\dd r} + \frac{\chi_{10}}{r}\right)\, d \phi=0.
\label{solv}
\end{equation}

\subsection{Law of motion}
Finally, to obtain the law of motion we substitute \eqref{firstinner}
into (\ref{solv}) to give
\begin{equation}
\frac{d {\bf X}_{\ell}}{dT}=-\frac{2  n_{\ell}}{\beta_{\ell}
  \tilde{\mu}}\, \del 
G^{\perp}({\bf X}_{\ell}),
\label{lawalfa1}
\end{equation}
where  $G$
is given by (\ref{G0eqn}).

\subsubsection{Law of motion for two spirals}

As an example consider two spirals at positions $(X_1,0)$ and
$(X_2,0)$, with $X_1<X_2$. In that case  
\[ \del G(X_1,0) = (-\beta_{2} \al K_0'(\al (X_2-X_1)),0).\]
 Using $\mu = 1/|\log \eps|$, $q = \pi/(2 |\log \eps|)$, 
and $\beta_{1} = \beta_{2}$  
 the law of motion reads
\beq
\frac{d {\bf X}_1}{dT}
= \left(0,\pi n_1\al K_0'(\al (X_2-X_1))\right).\label{2can}
\eeq
The direction in which each of a pair of spirals  moves depends only on
its own degree and not on the degree of the other spiral. This is
exactly the opposite to what happens in fluid dynamics vortices, which
would  
move depending only on the vorticity of the surrounding vortices.

If we now take the limit $q \ra 0$ we would expect the law of motion
(\ref{2can}) to tend to the law of motion (\ref{2q=0}) for $q=0$. Since we
scaled our outer region with $q$ by setting $q=\pi/(2 |\log \eps|)$,
letting $q\ra0$ is equivalent to letting $X_2-X_1 \ra 0$. With 
 $X_2-X_1 \ll 1$ equation (\ref{2can}) becomes
\beq
\frac{d {{\bf X}}_1}{dT}= \left(0,-\frac{n_1\pi}{ X_2- X_1}\right).
\label{2spirals}
\eeq
We see that this does not agree with (\ref{2q=0}); in fact, the two
velocities are orthogonal to each other.

In order to match the law for spirals at the canonical separation with
that for $q=0$ we need to consider an intermediate region, in which
the separation of spirals is much less than the outer core radius. We
analyse such a configuration in \S\ref{middle}. We will find that the
law of motion for such separations includes components both along and
perpendicular to the line of centres, and matches with both the laws
(\ref{2can}) and (\ref{2q=0}) in the appropriate limits.

\section{Interaction of spirals in the near-field}
\label{middle}
We now assume that the spirals are separated by  
distances smaller than the canonical separation.
The first step is to determine how this affects the eigenvalue
$\alpha$. If we expand (\ref{alphaK}) as $|{\bf X}_1-{\bf X}_2|\ra 0$
we find $\alpha \sim |{\bf X}_1-{\bf X}_2|^{-1/2}$ from which it seems
that $\alpha$ should be larger than order one in the present
scaling. However, we need to remember that we are redefining the
$\eps$ which appears in the definition of $\alpha$. 
Since $\alpha$ is inversely proportional to $\eps$, this redefinition
dominates the square root growth above, so that $\alpha$ is in fact
algebraically small in $\eps$ and 
exponentially small in $q$.

\subsection{Outer Region}
Proceeding as in \S\ref{canouter} and expanding
in powers of $\eps$ as 
\beqas
f &\sim& f_0(q,\m) + \e^2 f_1(q,\m) + \cdots\\
\ki&\sim &\ki_0(q,\m) +  \cdots,
\eeqas
we find
\beqa
 f_0 &=& 1,\qquad 
f_1 = -\frac{1}{2} |\n \ki_0|^2,\\
 \m \ki_{0T} &=& \n^2 \ki_0 + q|\n \ki_0|^2.\label{newchieqn}
\eeqa
As before we  write $\m=\tilde{\m}q$ and assume (which will be
justified {\em a posteriori}) that $\tilde{\mu}$ is of order one. 
Expanding
\[
\ki_0\sim \frac{\ki_{00}}{q}+\ki_{01}+q\ki_{02}+\cdots,
\]
 we find 
\beqa
0&=&\n^2 \ki_{00} + |\n \ki_{00}|^2,\label{one}\\
\tilde{\m}\ki_{00T}&=&\n^2\ki_{01}+2\n\ki_{00}\cdot\n\ki_{01},\label{two}
\eeqa
From \S\ref{sim} and \S\ref{canonical} we expect 
\begin{equation}
\ki_{00}=C_1(T)
\label{ki00out}
\end{equation}
to be constant in space,
while the topological singularities are contained in $\chi_{01}$, given by
\begin{equation}
\ki_{01}=C_{3}(T)+ \sum_{j=1}^N \left(n_j\phi_j+C_{2j}(T)\log
R_j+\frac{C_{1}'(T)}{4N}R_{j}^2\right), 
\label{ki01out}
\end{equation}
where $C_{2j}$ and $C_{3}$ are  time-dependent functions that will be
determined
by matching with the inner solution. As
before, we use the notation $R_j$ and $\phi_j$ to represent the polar
coordinates centred at the $j$-th spiral. 
Note that, since we are now working with $\chi$ rather than $h$, it is
 clear that the corresponding wavefunction $\psi$ is single valued. 

\subsection{Inner Region}
The expansion in the inner region proceeds exactly as in
\S\ref{caninner}.

\subsection{Asymptotic matching}
\subsubsection{Outer limit of the leading-order inner solution}

We again use the technique of  Sections \ref{sim} and \ref{canonical}
 which allows us to sum all the  terms in $q$ in  
the outer limit of the inner expansion. The outer limit of the
leading-order inner  
is again given by 
\beqas
f_0 &\sim& 1 -\frac{\eps^2}{2} |\n\widehat{\ki}_{00}|^2+\cdots,\\
\ki_0 &\sim& \widehat{\ki}_{00}(q)
+\cdots,
\eeqas
with
 $\widehat{\ki}_{00} = n_{\ell}\phi + (1/q)\log\widehat{H}_0(R)$
where,
\begin{equation}
\label{newsolutionH}
H_0 = A_{\ell}(q)\eps^{-iq n_\ell} R^{iq n_\ell}+B_{\ell}(q)\eps^{iq n_\ell} R^{-iq n_\ell},
\end{equation}
where 
\beqas
A_{\ell}(q) &\sim& \frac{1}{q} A_{\ell 0}+A_{\ell 1} +q A_{\ell 2} +\cdots,\\ 
B_{\ell}(q) &\sim& \frac{1}{q} B_{\ell 0}+B_{\ell 1} +q B_{\ell 2}
+\cdots.
\eeqas
As before, comparing with the inner expansion gives
\begin{align}
\label{newA0}
A_{\ell 0}-B_{\ell 0} =& 0,\\
\label{newA1}
\frac{(A_{\ell 1}-B_{\ell 1})}{A_{\ell 0}+B_{\ell 0}}i =& -c_{n_\ell}n_{\ell} \quad \textrm{for $k =
  1, \ldots, N$}. 
\end{align}
The remaining equations determining $A_{\ell}$ and $B_{\ell}$ will be found from
matching with the outer region.
\subsubsection{Outer limit of the first-order inner}
The outer limit of the first-order inner solution is identical to that
of \S\ref{4.3.4}, namely
\beqas
f_1 &\sim&  - \eps \del \widehat{\ki}_{00} \cdot \del
\widehat{\ki}_{10}+\cdots,\\
\ki_1 &\sim&
\frac{\widehat{\ki}_{10}(q)}{\eps}+\cdots,
\eeqas
where
\[ 
\widehat{\ki}_{10}  = \frac{\widehat{h}_1 e^{- q
    \widehat{\ki}_{00}}}{q }, 
\]
with
\beqa
\widehat{h}_1 &=& \mbox{}  -\frac{\mu  A_{\ell} \eps^{-i
    qn_{\ell}}(V_{1\ell}-i V_{2\ell})}{4  }
R^{ iqn_{\ell}
  +1}\, e^{(qn_{\ell}+i)\phi}
- \frac{\mu B_{\ell} \eps^{i qn_{\ell}}( V_{1\ell}+ i V_{2\ell})}{4  } R^{ -iqn_{\ell} +1}\, 
e^{(qn_{\ell}-i)\phi}
\nonumber \\  
&&\mbox{ }
+\gamma_1R^{1-iqn_{\ell}}\, e
^{(qn_{\ell}+i)\phi}
+\gamma_2R^{1+iqn_{\ell}}\, 
e^{(qn_{\ell}-i)\phi}.\label{5outinh1}
\eeqa

\subsection{Inner limit of the outer}
To compute the inner limit of the 
outer we rewrite solutions \eqref{ki00out} and \eqref{ki01out} in
terms of the inner variable ${\bf X} = {\bf 
  X}_{\ell}+\e {\bf x}$ or equivalently $R_{\ell}=|{\bf X}_{\ell}-{\bf
  X}|=\eps r$, and expand in powers of $\e$. This gives 
\beqa
\ki_0&\sim& \frac{C_1}{q}+
C_{3}+n_{\ell}\phi + C_{2\ell}\log (\eps r) +G({\bf X}_{\ell})+\e\n G({\bf X}_{\ell})\cdot{\bf
  x}+\ordre(\e^2),\label{outl}
\eeqa
where
\beq
G({\bf X}) = \sum_{j\neq \ell}^{N} n_j
\phi_{j}+C_{2j}\log |{\bf X}-{\bf X}_j|+
\frac{C_{1}'}{4N}|{\bf X}-{\bf X}_j|^2 ,\label{5G0}
\eeq
We can now match this inner limit of the outer with the outer limit of
the inner,  given by \eqref{newsolutionH}. Since
we have the full logarithmic expansion in the inner region, but only some
terms in the logarithmic expansion of the outer region, we must  write
both expansions in terms of the outer variable before comparing terms.

Expanding \eqref{newsolutionH} in powers of $q$ using the relation
$A_{\ell 0}=B_{\ell 0}$ gives
\begin{align}
\widehat{\ki}_{00}\sim& \frac{\log(A_{\ell
    0}(e^{-iqn_{\ell}\log\e}+e^{iqn_{\ell}\log\e})/q)}{q}+n_{\ell}\phi\nonumber\\ 
+&\frac{A_{\ell 1}e^{-iqn_{\ell}\log\e}+B_{\ell
    1}e^{iqn_{\ell}\log\e}}{A_{\ell
    0}(e^{-iqn_{\ell}\log\e}+e^{iqn_{\ell}\log\e})}+i
n_{\ell}\frac{e^{-iqn_{\ell}\log\e}-e^{iqn_{\ell}\log\e}}{e^{-iqn_{\ell}\log\e}+e^{iqn_{\ell}\log\e}}\log
R\label{inl}
\end{align}
\subsubsection{Leading order matching (1ti)(1to)=(1to)(1ti).}
We can now match the leading order terms by comparing \eqref{inl} with
the leading order terms in $\e$ in the expression for the outer
\eqref{outl}. 
We see that the leading term in (\ref{inl}) is $1/q \log 1/q$, which
is not present in (\ref{outl}), but which is a constant (and constants
are unimportant in $\chi$). Thus we see that the outer expansion
should really proceed as\footnote{this expansion can also be motivated
  by the single spiral solution (\ref{ki0sim}).}
\beqa
\ki_0&\sim& -\frac{\log q}{q} + \frac{C_1}{q}+C_{3}+n_{\ell}\phi +
C_{2\ell}\log (\eps r) + 
G({\bf X}_{\ell})+\e\n G({\bf X}_{\ell})\cdot{\bf
  x}+\cdots.\qquad\label{outla}
\eeqa
Then, matching (\ref{outla}) and (\ref{inl}) gives
\beqa
A_{\ell 0}(e^{-iqn_{\ell}\log\e}+e^{iqn_{\ell}\log\e})&=&
e^{C_1},\label{condicio1}\\ 
\frac{A_{\ell 1}e^{-iqn_{\ell}\log\e}+B_{\ell
    1}e^{iqn_{\ell}\log\e}}{A_{\ell
    0}(e^{-iqn_{\ell}\log\e}+e^{iqn_{\ell}\log\e})}&=&
C_3+G({\bf X}_{\ell}),\\ 
i
n_{\ell}\frac{e^{-iqn_{\ell}\log\e}-e^{iqn_{\ell}\log\e}}{e^{-iqn_{\ell}\log\e}+e^{iqn_{\ell}\log\e}}&=&C_{2\ell}. \label{C2}
\eeqa
\subsubsection{First order matching (2ti)(1to)=(1to)(2ti).}
We now match 
the O($\e$)  terms in \eqref{outl}  with the $O(\eps)$ terms in the 
outer limit of the leading-order inner, which are 
\beqas
\widehat{\ki}_{10}&=&-\frac{\m}{4q}\frac{A_{\ell}
  e^{-iqn_{\ell}\log\e}(V_{1\ell}-iV_{2\ell})R^{1+iqn_\ell}
e^{i\phi}}{\left(A_{\ell} R^{iqn_\ell}e^{-iqn_{\ell}\log\e}+B_{\ell}
  R^{-iqn_\ell}e^{iqn_{\ell}\log\e}\right)}\nonumber\\  
&&\qquad\mbox{ }-\frac{\m}{4q}\frac{B_{\ell} 
  e^{iqn_{\ell}\log\e}(V_{1\ell}+iV_{2\ell})R^{1-iqn_\ell}e^{-i\phi}}
{\left(A_{\ell} R^{iqn_\ell}e^{-iqn_{\ell}\log\e}+B_{\ell}
 R^{-iqn_\ell}e^{iqn_{\ell}\log\e}\right)}\\
&&\qquad\qquad\mbox{ }
+\frac{\gamma_1 R^{1-iqn} e^{i \phi} + \gamma_2 R^{1+iqn}
  e^{-i\phi}}{q\left(A_{\ell} R^{iqn_\ell}e^{-iqn_{\ell}\log\e}+B_{\ell}
 R^{-iqn_\ell}e^{iqn_{\ell}\log\e}\right)}
.\label{innerm}  
\eeqas
Expanding as $q \ra 0$,
\beqas
\widehat{\ki}_{10}&\sim&\left(\frac{4\gamma_{10}-\tilde{\m}A_{\ell0} 
  e^{-iqn_{\ell}\log\e}(V_{1\ell0}-iV_{2\ell0})}{4\left(A_{\ell0}
  e^{-iqn_{\ell}\log\e}+B_{\ell0} 
  e^{iqn_{\ell}\log\e}\right)}\right)R
e^{i\phi}\nonumber\\  
&&\qquad\mbox{ }+\left(\frac{4 \gamma_{20}-\tilde{\m}B_{\ell0} 
  e^{iqn_{\ell}\log\e}(V_{1\ell0}+iV_{2\ell0})}
{4\left(A_{\ell0}e^{-iqn_{\ell}\log\e}+B_{\ell0}
 e^{iqn_{\ell}\log\e}\right)}\right)Re^{-i\phi}.
\eeqas
This must match with ${\bf X}\cdot \del G({\bf X}_\ell)$, giving
\beqas
 \gamma_{10} &=& \frac{\tilde{\mu}  A_{\ell0}  e^{-iqn_{\ell}\log\e} 
(V_{1\ell0}-i V_{2\ell0})}{4  }+\frac{G_{X}({\bf X}_{\ell}) - i G_{Y}({\bf
    X}_{\ell})}{2}\left(A_{\ell0}e^{-iqn_{\ell}\log\e}+B_{\ell0}
 e^{iqn_{\ell}\log\e}\right), \\
\gamma_{20} &=& \frac{\tilde{\mu} B_{0\ell}  e^{iqn_{\ell}\log\e}
( V_{1\ell0}+ i V_{2\ell0})}{4  } +\frac{G_{X}({\bf X}_{\ell}) + i G_{Y}({\bf
    X}_{\ell})}{2}\left(A_{\ell0}e^{-iqn_{\ell}\log\e}+B_{\ell0}
 e^{iqn_{\ell}\log\e}\right).
\eeqas
Finally we are ready to write $\widehat{\chi}_{10}$ back in terms of
the inner variable $r$ and expand in $q$ to give 
\begin{eqnarray}
\chi_{10} & \sim&-\frac{\tilde{\m}r}{4}\left(V_{1\ell0}\cos \phi + V_{2\ell0}\sin\phi\right)
+\frac{\tilde{\m}
  r}{4}\left(V_{1\ell0}\cos(\phi-2qn_{\ell}\log\e)+V_{2\ell0}\sin(\phi-2qn_{\ell}\log\e)\right)\nonumber\\ 
&&+r\cos(qn_{\ell}\log\e)\big(G_{X}({\bf
  X}_{\ell})\cos(\phi-qn_{\ell}\log 
    \e)+G_{Y}({\bf 
    X}_{\ell})\sin(\phi-qn_{\ell}\log \e)\big).
\label{ji1r}
\end{eqnarray}

\subsection{Law of motion}
The solvability condition on the first-order inner proceeds exactly as
in \S\ref{sim}. To obtain the law of motion we substitute 
\eqref{ji1r} into \eqref{solv} to find
\begin{equation}
\frac{d{\bf X}_{0 \ell}}{dT}=\frac{2}{\tilde{\mu}}
\frac{\cos(qn_{\ell}\log\e)}{\sin(qn_{\ell}\log\e)}\n G({\bf X}_{\ell})^{\perp}.
\label{lawalsmall}
\end{equation}
Differentiating (\ref{5G0}) gives
\[\del G({\bf X}) = \sum_{j\neq \ell}^{N}
\left(
n_j\frac{{\bf e}_{\phi_j}}{|{\bf X}_j-{\bf X}_{\ell}|}
+C_{2j}\frac{{\bf e}_{r_j}}{|{\bf X}_j-{\bf X}_{\ell}|}
\right).
 \]
Using (\ref{C2}) this becomes
\begin{equation}
\frac{d{\bf
    X}_\ell}{dT}\sim\frac{2}{\tilde{\m}}
\frac{\cos(qn_{\ell}|\log\e|)}{\sin(qn_{\ell}|\log\e|)}\sum_{j\neq
    \ell}^{N} 
\left(n_j\frac{{\bf e}_{r_j}}
{|{\bf X}_j-{\bf
    X}_{\ell}|}+n_j\frac{\sin(qn_j|\log\e|)}{\cos(qn_j|\log\e|)}
\frac{{\bf e}_{\phi_j}}{|{\bf X}_j-{\bf X}_{\ell}|}\right). 
\label{lawalsmall2}
\end{equation}
Since $n_j = \pm 1$ for all $j$, we can simplyfy (\ref{lawalsmall2}) to
\begin{equation}
\frac{d{\bf X}_\ell}{dT}\sim 2n_{\ell}q|\log \eps|\sum_{j\neq
  \ell}^{N} 
\left(n_j\frac{\cos(q|\log\e|)}{\sin(q|\log\e|)}
\frac{{\bf e}_{r_j}}
{|{\bf X}_j-{\bf X}_{\ell}|}+
\frac{{\bf e}_{\phi_j}}{|{\bf X}_j-{\bf X}_{\ell}|}\right),\label{middlelaw}
\end{equation}
where we have taken $\mu = 1/|\log \eps|$ as usual.
We see that the law of motion in the near-field region interpolates
between motion along the line of centres and motion perpendicular to
the line of centres as $q |\log \eps|$ varies from $0$ to $\pi/2$.

\subsubsection{Matching with the other limits} 
\label{lawmatch}
If we  take the limit as $q|\log \eps|\ra0$ in (\ref{middlelaw}) we find
\[
\frac{d{\bf X}_\ell}{dT}\sim 2n_{\ell}\sum_{j\neq
  \ell}^{N} 
\frac{n_j{\bf e}_{r_j}}
{|{\bf X}_j-{\bf X}_{\ell}|},
\]
in agreement with \eqref{lawq0}.  

If we  take the limit as $q|\log \eps|\ra\pi/2$ in (\ref{middlelaw}) we find
\beq
\frac{d{\bf X}_\ell}{dT}\sim n_{\ell}\pi\sum_{j\neq
  \ell}^{N} 
\frac{{\bf e}_{\phi_j}}{|{\bf X}_j-{\bf X}_{\ell}|}.\label{midlim}
\eeq
This should match with the limit $\nu \ra -\infty$ of
(\ref{lawalfa1}).
As $\nu \ra -\infty$ we see from (\ref{alphaK}) that  $\alpha \ra 0$,
so that
\beqas
\del G({\bf X}) &\sim& -\sum_{j=1,j\neq\ell}^{N} \beta_{j}(T)
\frac{{\bf e}_{r_j}}{|{\bf
  X}-{\bf X}_j|}.
\eeqas
Now, letting $\nu\ra-\infty$, $\al \ra 0$ in (\ref{90}) 
 gives, at leading order,
\[
-\nu \beta_{\ell} =  - 
\sum_{j=1}^{N}\beta_{j} \log \alpha .
\]
with solution $\alpha = e^{\nu/N}$, $\beta_{j} = \beta_0
= $ constant, for all $j$, in agreement with (\ref{alphaK}). 
Thus in the limit as $\nu \ra -\infty$ (\ref{lawalfa1}) gives
\[
\frac{d {\bf X}_{\ell}}{dT}\sim
\pi  n_{\ell} \sum_{j=1,j\neq\ell}^{N} 
\frac{{\bf e}_{\phi_j}}{|{\bf
  X_{\ell}}-{\bf X}_j|},
\]
in agreement with (\ref{midlim}).

\section{Conclusions}
\label{numerics}
The main contribution in this paper is the description of complicated
patterns of the complex Ginzburg-Landau equation with many moving
spirals in terms of simple sets of ordinary differential equations
that provide a law of motion for the centres of the spirals. We have
focussed on the case of spirals with winding number $\pm 1$ (one-armed
spirals). Our 
results rely on the structural stability of such spirals; this 
is discussed in \cite{hagan82} where it is conjectured that 
$1$-armed spirals are the most stable.  

Our analysis is based on the limit $q \ra 0$, where $q$ is a measure
of  the imaginary component of the coefficients in the equation. 
We find that the cases $q=0$ and $q>0$ behave very differently. For
$q>0$ each spiral has an inner core (in which the magnitude of the
wavefunction varies from zero to one), and an outer core (in which the
level lines of the phase of the wavefunction vary from radial to
azimuthal). The radius of the outer core can be related to the
asymptotic wavenumber (the wavenumber at infinity) for a single spiral, and is
exponentially large in $q$, tending to infinity as $q \ra 0$.
We find that the law of motion of spirals depends on the relative
sizes of the separation and the outer core radius. Furthermore, the
whole pattern oscillates at a frequency which varies slowly as the
spirals move.
 Our main results can be summarised as follows.

\subsection{Laws of motion} 
\begin{figure}
\begin{center}
\input{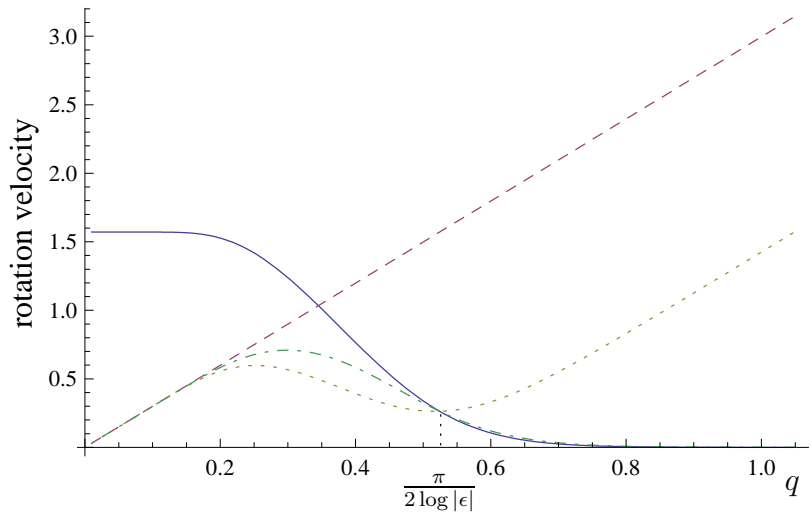}
\end{center}
\caption{The velocity of rotation of a pair of $+1$ spirals as a function
  of $q$ for $\eps = 1/20$ and $|{\bf X}_1 - {\bf X}_2|=2$ using the
  formulae for the canonical scaling (solid), the near-field scaling
  (dashed), the additative composite (dotted) and the multiplicative
  composite (dot-dashed). We see that the multiplicative composite is
  likely to be a much better approximation, especially for
  $q>\pi/2\log|\eps|$.}
\label{fig:comp}
\end{figure}

For $0<q\ll1$, given a set of $\pm 1$-armed
spirals with mutual separation of order $1/\e$, the spirals evolve 
on a time-scale given by $T= \e^2
t/|\log\e|$ and satisfy the following laws of motion: 
\begin{enumerate}[(i)] 
\item In the so-called \emph{canonical separation} of the spirals,
  which corresponds to \linebreak\mbox{$q|\log\e|=\pi/2+q\nu+\ordre(q^2)$}, with
  $\nu=\ordre(1)$, the spirals' centres satisfy the law of motion 
\beqa
\frac{d {\bf X}_{\ell}}{dT}&\sim&
-\frac{\pi n_{\ell}\alpha}{\beta_{\ell} }\,\sum_{j\neq\ell}^{N}
\beta_{j} K_0'(\alpha |{\bf X}_{\ell}-{\bf
  X}_j|)\,{\bf e}_{\phi_j}, 
\label{canfin}
\eeqa
where $c_1=0.098...$, along with the set of linear equations
\begin{equation}
0 = \beta_{\ell}(c_{1}+\nu-\log\alpha+\log 2-\gamma) + \sum_{j\neq\ell}^{N}\beta_{j} K_0(\alpha |{\bf X}_{\ell}-{\bf X}_j|),
\label{wavenumber}
\end{equation}
for the parameters $\beta_{j}$, whose solvability condition
determines $\alpha$.  This solvability condition shows that the
corresponding asymptotic wavenumber, $k=\eps \al/q$, 
 is exponentially small in $q$ and evolves in time along
with the positions of the spirals. 
\item In the so-called \emph{near-field separation} of the spirals, which
  corresponds to inter-spirals separations such that
  $0<q|\log\e|<\pi/2$, the spirals' centres  
satisfy the law of motion
\beqa
\frac{d{\bf X}_{\ell}}{dT} &\sim& 2n_{\ell}q\log|\eps|\sum_{j\neq
  \ell}^{N} \left(n_j\frac{\cos(q|\log\e|)}{\sin(q|\log\e|)}\frac{{\bf
    e}_{r_j}}{|{\bf X}_j-{\bf X}_{\ell}|}+\frac{{\bf e}_{\phi_j}}
{|{\bf X}_j-{\bf X}_{\ell}|}\right).
\label{midfin}
\eeqa
\item 
In the limit as $q\to 0$ the law of motion (\ref{midfin}) agrees with the one
obtained by Neu in \cite{neu90}, 
\begin{equation}
\frac{d{\bf X}_{\ell}}{dT} \sim 2n_{\ell}\sum_{j\neq {\ell}}
\frac{n_j{\bf e}_{r_{j}}}{|{\bf X}_{\ell}-{\bf X}_j|}.
\end{equation}
\end{enumerate}
We can use the laws of motion in (i) and (ii) to form a composite
expansion valid throughout the range $0<q\log|\eps|\leq \pi/2$. We saw
in \S\ref{lawmatch} that the law of motion in the overlap region
(i.e. the limit of (\ref{canfin}) as $\nu \ra -\infty$, which is the limit of
(\ref{midfin}) as $q\log|\eps| \ra \pi/2$) is (\ref{midlim}). Adding
(\ref{canfin}) and (\ref{midfin}) and subtracting (\ref{midlim}) gives
the (additative) composite law of motion as
\beqa
\frac{d {\bf X}_{\ell}}{dT}&\sim&
-\frac{\pi n_{\ell}\alpha}{\beta_{\ell} }\,\sum_{j\neq\ell}^{N}
\beta_{j} K_0'(\alpha |{\bf X}_{\ell}-{\bf
  X}_j|)\,{\bf e}_{\phi_j}\nonumber\\
&&\mbox{ }+2n_{\ell}q\log|\eps|\sum_{j\neq
  \ell}^{N} \left(n_j\frac{\cos(q|\log\e|)}{\sin(q|\log\e|)}\frac{{\bf
    e}_{r_j}}{|{\bf X}_j-{\bf X}_{\ell}|}+\frac{{\bf e}_{\phi_j}}
{|{\bf X}_j-{\bf X}_{\ell}|}\right)\nonumber\\
&&\mbox{ }-
\pi  n_{\ell} \sum_{j=1,j\neq\ell}^{N} 
\frac{{\bf e}_{\phi_j}}{|{\bf
  X_{\ell}}-{\bf X}_j|}.
\eeqa
Alternatively, we can form the (multiplicative) composite expansion 
\beqa
\frac{d {\bf
    X}_{\ell}}{dT}&\sim&-\frac{2n_{\ell}q\log|\eps|\alpha}{\beta_{\ell
    0} }\sum_{j\neq 
  \ell}^{N}
\beta_{j}K_0'(\alpha |{\bf X}_{\ell}-{\bf  X}_j|)
\left(n_j\frac{ \cos(q|\log\e|)}{\sin(q|\log\e|)}\,{\bf 
    e}_{r_j}+{\bf e}_{\phi_j}\right)\label{comp2}
\eeqa
which has the correct asymptotic limit in each region.
These formulae are illustrated in Figure \ref{fig:comp}. The
additative composite is accurate in the region it was constructed to
hold ($q \log |\eps| \leq \pi/2$), but the multiplicative composite
remains a good approximation even for $q \log |\eps| > \pi/2$.

We see that for small separations the interaction is along the line of
centres, with like spirals repelling, opposites attracting. However,
as the separation increases the direction of the interaction gradually
changes, until it is perpendicular to the line of centres at large
distances.

While the attraction/repulsion of spirals depends on the winding
numbers of both,
the direction of rotation of one spiral about another depends only on
its own winding number: the centre of positive spirals rotates around
any other wpiral in an
anti-clockwise direction, while the centre of a
negative spiral rotates in a clockwise direction about another spiral.
Thus like positive spirals rotate in an anti-clockwise direction while
separating, like negative spirals rotate in an anti-clockwise
direction while separating, and unlike spirals translate while
approaching.

Since the motion is perpendicular to the line of centres in the
canonical scaling, the question arises as to the existence of bound
states. To answer this question the first-order correction to the
radial velocity is needed. Our calculations indicate that the radial
velocity remains of one sign, so that bound states are not possible
for small $q$, in agreement with \cite{aranson93}, who found that bound states
are only possible for $q>0.845$.

\subsection{Comparison with direct numerical simulations}

To compare with numerical simulations we write (\ref{comp2}) in terms of
the original variables ${\bf x}$ and $t$ to give
\beqa
\frac{d {\bf
    x}_{\ell}}{dt}&\sim&-\frac{2n_{\ell}q\hat{\alpha}}{\beta_{\ell
    0} }\sum_{j\neq 
  \ell}^{N}
\beta_{j}K_0'(\hat{\alpha}|{\bf x}_{\ell}-{\bf  x}_j|)
\left(n_j\frac{ \cos(q|\log\e|)}{\sin(q|\log\e|)}\,{\bf 
    e}_{r_j}+{\bf e}_{\phi_j}\right) \label{comp}
\eeqa
where $\hat{\alpha} = \eps \alpha$ satisfies
\begin{equation}
0 = \beta_{\ell}\left(c_{1} - \frac{\pi}{2q}-\log\hat{\alpha}+\log
  2-\gamma\right) + 
\sum_{j\neq\ell}^{N}\beta_{j} K_0(\hat{\alpha} |{\bf x}_{\ell}-{\bf x}_j|).
\label{wavenumber2}
\end{equation}

\begin{figure}
\begin{center}
\includegraphics[clip,width=12cm]{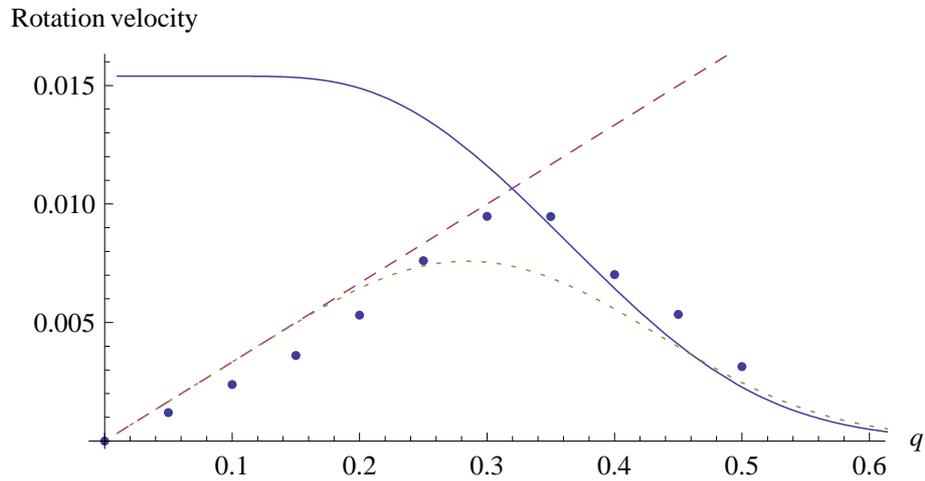}
\caption{Rotation velocity of a pair of spirals 60 units apart as a
  function of $q$, using (\ref{canfin}) [solid], (\ref{midfin})
  [dashed] and the composite (\ref{comp2}) [dotted]. The points
  correspond to a   numerical simulation of (\ref{CGL0}). }
\label{fig1}
\end{center}
\end{figure}

\begin{figure}
\begin{center}
\includegraphics[clip,width=12cm]{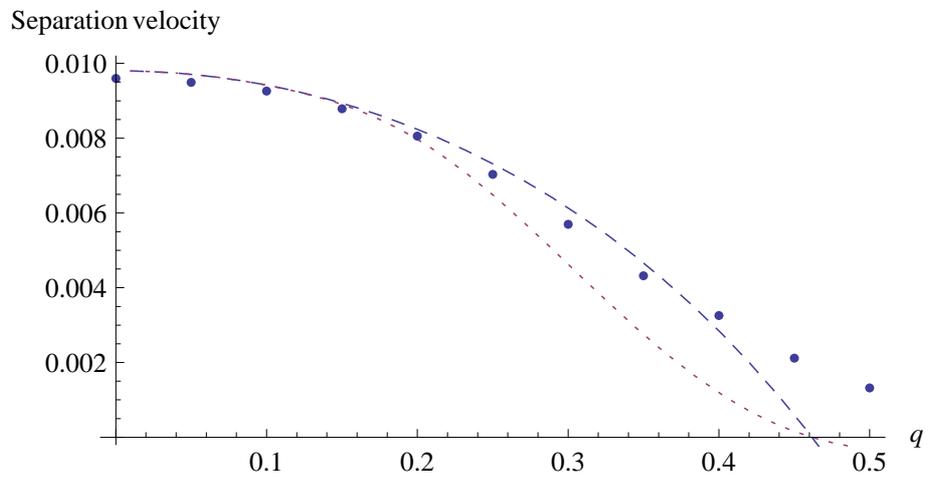}
\caption{Separation velocity of a pair of spirals 60 units apart using
  (\ref{midfin})   [dashed] and the composite (\ref{comp2}) [dotted].}
\label{fig2}
\end{center}
\end{figure}

In Figures \ref{fig1} and \ref{fig2} we compare the predictions
of the canonical, near-field and composite laws of motion 
with a direct simulation of (\ref{CGL0}) using second-order accurate
finite differences in a square domain of side 800. The figures show
the separation and rotation velocity of a pair of $n_{\ell}=1$ spirals
when they are separated by a distance 60.

We see that the multiplicative composite expansion captures the
qualitative behaviour very well, and provides a reasonable
quantitative prediction.

\paragraph{Acknowledgements} M. Aguareles was supported in part by the
MEC of Spain, grant MT2005-07660-C02-01. 
The authors would like to acknowledge many helpful discussions with
Prof T Witelski, who performed the numerical simulations reported in
\S\ref{numerics}. 

\def\cprime{$'$}

\appendix

\section{Extension to non-zero $b$}
For clarity throughout we have only considered the
case $b=0$. However,  the extension to general values of $b$ does not
introduce any further conceptual difficulty, it merely complicates some
of the algebra. We indicate here briefly the necessary steps in the
calculation.

The outer equation for nonzero $b$ reads
\begin{equation}
\e^2(1-ib)\m\p_T = \e^2 \n^2\p+(1+iq)(1-|\p|^2)\p-\frac{i\e^2\alpha^2}{q}\p\label{outeral1},
\end{equation}
which in terms of modulus $f$ and phase $\ki$ becomes
\begin{align}
\m \e^2 (f_T+bf\ki_T) &= \e^2 \n^{2} f -\e^2 f |\n \ki|^2+ (1-f^2)f,\\
\m \e^2(f^2 \ki_T-bf f_T)&= \e^2\n\cdot (f^2\n\ki)+qf^2(1-f^2)-\e^2\frac{\alpha^2}{q}f^2.
\end{align}
Upon expanding $\ki$ and $f$ in powers of $\e$  we find that the
leading-order terms satisfy 
\begin{align*}
f_0 &= 1,\\
\m \ki_{0T} &= \n^2 \ki_0 + q|\n\ki_0|^ 2-\al^2/q.
\end{align*}
where the parameter $b$ does not appear; thus the rest of the outer
calculation is identical to that of \S\ref{canouter}. 

For $b \not = 0$, the inner equations close to the $\ell$-th vortex 
 are given by 
\begin{align}
\e\m \big(\e f_T-\frac{d{\bf X}_{\ell} }{dT}\cdot \n f &+ \e bf\ki_T-b f\frac{d{\bf X}_{\ell} }{dT}\cdot \n \ki\big),\nonumber\\
&=\n^2 f -f|\n\ki|^2+(1-f^2)f\\
\e\m \big(b\frac{d{\bf X}_{\ell} }{dT}\cdot f\n f-\e b f f_T &+ \e  f^2\ki_T- f^2\frac{d{\bf X}_{\ell} }{dT}\cdot \n \ki\big)\nonumber\\
&= \n\cdot(f^2\n\ki)+q(1-f^2)f-\frac{\e^2\al^2}{q}.
\end{align}
Expanding in powers of $\e$ we find that at leading order
\begin{equation}
0=\n^2 \psi_0+(1+iq)\psi_0(1-|\psi_0|^2),
\end{equation}
while at first order
\begin{equation}
-\m(1-ib)\frac{d{\bf X}_{\ell}}{dT} \cdot \n \p_0
=\n^2\p_1+(1+iq)(\p_1(1-|\p_0|^2)-\p_0(\p_0\p_1^*+\p_0^*\p_1)), 
\label{1in5} 
\end{equation}
or equivalently,
\beqa
- \m \frac{d{\bf X}_{\ell}}{dT}\cdot(\n f_0+bf_0\n\ki_0) &=& \n^2 f_1 
-f_1|\n  \ki_0|^2-2 f_0 \n  \ki_0 \cdot \n \chi_1
+f_1 - 3 f_0^2 f_1,\\
- \m\frac{d{\bf X}_{\ell}}{dT}\cdot( f_0^2\n \ki_0-bf_0\n f_0) &=&\n \cdot(f_0^2\n
  \ki_1)+\n \cdot(2 f_0 f_1\n \ki_0)
+2 q f_0 f_1 -4 q f_0^3 f_1.\qquad
\eeqa
Writing these in terms of the outer variable to find the outer limit of the
first order inner, and expanding in $\e$ as $\ki_1 \sim
\widehat{\ki}_{10}(q)/\eps +\widehat{\ki}_{11}(q)+\cdots$ and 
$f_1 \sim \widehat{f}_{10}(q) +\e\widehat{f}_{11}(q)+\cdots$ as in
\S\ref{4.3.4}, we find 
the equation for $\widehat{\ki}_{10}$ is now given by
\begin{equation}
- \m  \frac{d{\bf X}_{\ell}}{dT}\cdot\n \widehat{\ki}_{00} =  
  \n^2  \widehat{\ki}_{10} -2 q  \widehat{f}_{11} = 
\n^2 \widehat{\ki}_{10}+ 2 q \del \widehat{\ki}_{00} \cdot \del
\widehat{\ki}_{10},
\end{equation}
which again does not involve $b$.
Thus the matching between inner and outer solutions is the same as in
\S\ref{canonical}.

Finally we must derive the law of motion when $b$ is not zero. To do
so we start by defining the same inner product as before and consider
the first order linear operator given in \eqref{1in5}, which is of the
form 
$$L(q,\p_0)[\p_1] = w(\p_0,q,\m,b,d{\bf X}_{\ell}/dT)$$ 
where $L$ does not depend on $b$ but where the non-homogeneous term is now 
\begin{equation*}
w(\p_0,q,\m,b,d{\bf X}_{\ell}/dT) = -\m(1-ib) \frac{d{\bf X}_{\ell}}{dT} \cdot \n \p_0.
\end{equation*}
Since the linear operator $L$ is the same that that in
\S\ref{canonical},
 the only place where $b$ has an effect is in the solvability
 condition itself, which  is obtained through the Fredholm
 Alternative. The new solvability condition is 
\beq
-\int_D \Re\left\{\m(1-ib)\frac{d{\bf X}_{\ell}}{dT}\cdot \n \p_0 v^*\right\}
dD
=\int_{\partial
  D}\Re\left\{(1-iq)\left(v^*\frac{\partial \p_1}{\partial n}-\frac{\partial
  v^*}{\partial n}\p_1\right)\right\} dl. 
\label{solvability5}
\eeq
Since the left-hand side is $O(\mu)$, a nonzero value of $b$ does not
affect the velocity law at leading order, but it will alter the
$O(\mu)$ correction terms.

\end{document}